\documentclass[sigconf]{acmart}

\newcommand\vldbdoi{XX.XX/XXX.XX}
\newcommand\vldbpages{XXX-XXX}
\newcommand\vldbvolume{14}
\newcommand\vldbissue{1}
\newcommand\vldbyear{2020}
\newcommand\vldbauthors{\authors}
\newcommand\vldbtitle{\shorttitle} 
\newcommand\vldbavailabilityurl{https://github.com/LiuXiaoxuanPKU/ConstrOpt}
\newcommand\vldbpagestyle{plain} 
\usepackage{enumerate}
\usepackage{graphicx}
\usepackage{float}
\newfloat{algorithm}{t}{lop}
\usepackage{subfigure}
\usepackage{color}
\usepackage{tabularx}
\usepackage{hyperref}
\usepackage{ragged2e}  
\usepackage{booktabs}
\usepackage{algorithm}
\usepackage{algorithmicx}
\usepackage{adjustbox}
\usepackage{tikz}
\usepackage{multirow}
\usepackage[noend]{algpseudocode}
\usepackage{fixltx2e}
\usepackage{caption} 
\usepackage{comment}
\usepackage{hhline}
\usepackage[labelfont=bf]{caption}
\usepackage{etoolbox} 
\usepackage{mdwlist} 
\usepackage[inline]{aplcomments}
\usepackage{xspace}
\usepackage{paralist}
\usepackage{listings}
\usepackage{tcolorbox}
\usepackage{color,calc}
\usepackage{bold-extra}
\usepackage{enumitem}
\usepackage{graphics}
\usepackage{balance}

\newcommand{\tool}{\textsc{ConstrOpt}\xspace}
\newcommand{\code}[1]{\tt{#1}}

\newcommand{\seclabel}[1]{\label{#1}}

\newcommand{\secref}[1]{Sec.~\ref{#1}}

\newcommand{\figref}[1]{Figure~\ref{#1}}
\newcommand{\tabref}[1]{Table~\ref{#1}}
\newcommand{\lstref}[1]{Listing~\ref{#1}}

\newcommenter{shan}{1.0,0.0,0.0}
\newcommenter{junwen}{1.0,0.7,0.0}
\newcommenter{lily}{1.0, 0.5, 0.0}
\newcommenter{alvin}{0.2,0.8,1.0}
\newcommenter{zoey}{0.3,0.7,0.0}
\newcommenter{donny}{0.5,0.5,0.8}
\newcommand{\revision}[1]{\textcolor{brown}{#1}}

\setcopyright{none} 
\renewcommand\footnotetextcopyrightpermission[1]{}

\begin{document}
\pagestyle{plain}
\title{Leveraging Application Data Constraints to Optimize Database-Backed Web Applications}

\author{Xiaoxuan Liu$^\dagger$, Shuxian Wang$^\dagger$, Mengzhu Sun$^\dagger$, Sicheng Pan$^\dagger$, Ge Li$^\dagger$, Siddharth Jha$^\dagger$, Cong Yan$^\mathsection$, Junwen Yang$^\ddagger$, Shan Lu$^\mathparagraph$, Alvin Cheung$^\dagger$}

\affiliation{\emph{$^\dagger$UC Berkeley, $^\mathsection$Microsoft Research, $^\mathparagraph$University of Chicago, $^\ddagger$Meta}}

\begin{abstract}
Exploiting the relationships among data is a classical query optimization technique. 
As persistent data is increasingly being created and maintained programmatically, 
prior work that infers data relationships from data statistics misses an important opportunity.
We present \tool, the first tool that identifies data relationships by analyzing database-backed applications.
Once identified, \tool leverages the constraints to optimize the application's physical design
and query execution. 
Instead of developing a fixed set of predefined rewriting rules, \tool employs an enumerate-test-verify  
technique to automatically exploit the discovered data constraints to improve query execution.
Each resulting rewrite is provably equivalent to the original query.
Using 14 real-world web applications, our experiments show that 
\tool can discover numerous data constraints from code analysis and improve real-world application performance significantly.
\end{abstract}
\maketitle

\pagestyle{\vldbpagestyle}
\begingroup\small\noindent
\raggedright\textbf{PVLDB Reference Format:}\\
\vldbauthors. \vldbtitle. PVLDB, \vldbvolume(\vldbissue): \vldbpages, \vldbyear.\\
\href{https://doi.org/\vldbdoi}{doi:\vldbdoi}
\endgroup
\begingroup
\renewcommand\thefootnote{}\footnote{\noindent
This work is licensed under the Creative Commons BY-NC-ND 4.0 International License. Visit \url{https://creativecommons.org/licenses/by-nc-nd/4.0/} to view a copy of this license. For any use beyond those covered by this license, obtain permission by emailing \href{mailto:info@vldb.org}{info@vldb.org}. Copyright is held by the owner/author(s). Publication rights licensed to the VLDB Endowment. \\
\raggedright Proceedings of the VLDB Endowment, Vol. \vldbvolume, No. \vldbissue\ %
ISSN 2150-8097. \\
\href{https://doi.org/\vldbdoi}{doi:\vldbdoi} \\
}\addtocounter{footnote}{-1}\endgroup

\ifdefempty{\vldbavailabilityurl}{}{
\vspace{.3cm}
\begingroup\small\noindent
\raggedright\textbf{PVLDB Artifact Availability:}\\
The source code, data, and/or other artifacts have been made available at \url{\vldbavailabilityurl}.
}
\endgroup

\section{Introduction} 
From key constraints to the uniqueness of data values, relationships among attributes in a dataset are bases for relational query optimization. 
These {\em data constraints} occur in datasets across many different application domains~\cite{yang:icse20, li2016wander}, 
and have been used in optimization that ranges from normalizing relational schemas~\cite{bcnf}, 
detecting data errors~\cite{jiannan:sigmod20:error-detection}, 
to leveraging functional dependencies to improve query execution~\cite{FD-QO-theory}. 

There has been a long line of research that applies statistical~\cite{huhtala1999tane} and machine learning~\cite{de2018learning} techniques to identify data constraints from persistently stored data. While such data-driven approaches have been effective in discovering constraints from {\em collected} datasets (such as those from census or physical experiments), we are unaware of techniques that target {\em programmatically-generated} datasets. 

We encounter programmatically-generated datasets routinely in our daily lives---all websites process user inputs via web applications that generate persistently stored data. While there are means to express data constraints for programmatically-generated data, such as SQL constraints~\cite{cochrane1996integrating} and various {\em data validation} APIs provided by web application frameworks~\cite{django-validation, rails-validation}, they all require developers to manually declare them in their applications, which has shown to be tedious and error-prone to developers due to their complexity~\cite{yang:icse20}. As the artifacts that are used to generate or manipulate such datasets are often available (e.g., web application code, synthetic data generators), relying on data analysis techniques to ``re-discover'' constraints by analyzing the stored data while ignoring the programmatic artifacts is simply a missed opportunity.

In this paper, we investigate the feasibility of discovering data constraints by analyzing the programs that generate and process stored data. We focus on database-backed web applications, given their prevalence and the public availability of artifacts.
We study how such applications are developed and designed \tool, the first tool that automatically analyzes database-backed web applications to discover different data constraints in the stored data. Given the source code of a web application, \tool statically analyzes the code to extract candidate constraints.

We then demonstrate the usefulness of the discovered constraints by using them to optimize application performance. 
\tool leverages the extracted constraints to change the data schema to reduce storage, 
add parameter precheck to avoid issuing queries, and rewrite queries to improve their performance. 
To optimize query performance, we first install the extracted constraints into the database
and see if the optimizer can use those constraints to optimize queries.
As we discuss in~\secref{sec:queryrewrites}, 
mainstream commercial and open-source databases fail to utilize \tool-extracted constraints as they rely on pattern-matching rules to rewrite queries~\cite{finance1991rule, pirahesh1992extensible, begoli2018apache, blog-disjunct, join-elimination}, and those rules are often application-specific.
While we can modify the optimizer with additional rules, doing so can take substantial effort\footnote{It took more than two years for PostgreSQL developers to implement a pattern that removes the {\code DISTINCT} clause if the result is unique by definition, and this feature is yet to be merged~\cite{pg-remove-distinct}.} and might only benefit a few applications as rules are overly specific.

Moreover, as previous work has shown~\cite{cheng1999implementation}, blindly applying such rules can degrade query performance.
\tool instead uses an ``enumerate-test-verify'' approach, where a number of rewrites are first {\em enumerated} for each query based on various query features (e.g., whether it joins one of the tables with identified constraints). \tool subsequently estimates the cost of the rewritten query using the database's optimizer. If the cost is less, it verifies that the rewritten and original queries are semantically equivalent using both test cases and a formal verifier before deploying them.

We evaluate \tool by using it to discover and optimize widely-deployed database-backed web applications automatically. Results show that \tool can discover a wide variety of data constraints and \tool’s automatic query rewrite can detect 2511 rewrites with extracted constraints.

In sum, this paper makes the following contributions:
\begin{asparaitem}
    \item Data constraints are often embedded in artifacts that generate and process persistent data programmatically.
    To our knowledge, this is the first work that discovers data constraints from application source code and uses them for query optimization.

    \item We use the \tool-extracted constraints to automatically rewrite queries, optimize application code, and change the physical design (\secref{sec:changecodeschema}).
    To rewrite queries, rather than crafting a priori query transformation rule, we enumerate candidate query rewrites and utilize formal 
    verification to search for equivalent and efficient ones. This results in a simple yet general query optimizer that can significantly improve application performance (\secref{sec:queryrewrites}).
    
    \item 
    Evaluation using 14 popular open-source web applications shows that \tool can extract 4039 constraints (averaging 289 per application), 
    with few of them previously installed in the database by the application. 
    We evaluate \tool's optimization on 6 apps.
    Among queries with constraints, 13.8\% queries can benefit from data layout optimization, and 47\% queries are optimized by changing application code. 
    Finally, \tool’s constraint-driven optimizer improves the performance of 2511 queries, 118 of which has over 2$\times$ speedup.
\end{asparaitem}
\section{Background}
\seclabel{sec:background}
\subsection{Structure of ORM applications}
Applications built using the Object-relational mapping (ORM) framework are structured using the model-view-controller (MVC) architecture. 
For example, when a web user submits a form through a URL like
{\code http://foo.com/wikis/id=1}, a \emph{controller} action
{\code wikis/index} is triggered. This action takes the parameters
from the HTTP request (e.g., “{\code 1}” in the URL)
and interacts with the database by calling the Rails' ActiveRecord API, which  
translates the request into SQL queries
(e.g., a select query to retrieve wiki record with {\code id=1}). The results are then serialized into
\emph{model} objects (e.g., a {\code Wiki} object) and returned to the controller.
The returned objects are then passed to the \emph{view} files to generate
a webpage that is sent back to users. 

\subsection{Associations among classes and tables}
\seclabel{associations}
ORM frameworks provide an object-oriented interface to manage persistent data, where each class or class hierarchy is mapped to database table(s). 
To support inheritance in relational databases, ORM frameworks either use one table to store data for all types under the same inheritance 
hierarchy (also called the ``Table Per Hierarchy'' approach), or store each type in its separate table (i.e., ``Table Per Type''). 
Table Per Hierarchy results in one table storing all entities in the inheritance hierarchy. The table includes a ``discriminator'' column, which stores the actual type for each row.
Developers can define relationships
between classes, such as {\code belongs\_to}, {\code has\_one}, and {\code has\_many}. 
Once defined, an object can simply retrieve its relevant objects of different classes
through its fields without using joins.

\subsection{Data constraints in ORM applications}
Data constraints are rules enforced on stored data. Including class relationships, there are three ways to express data constraints:  

\noindent{\textbf{Front-end constraints.}} Developers can check for user input on the client (e.g., web browser). For instance, enforcing that password 
length must be longer than 6 characters and returns an error on failures without contacting the application server. 

\noindent{\textbf{Application constraints.}} The application code running on the server can also contain constraints. 
For instance, it can validate data values before inserting them into the database and only persist data when validation passes. 
As we will discuss in \secref{sec:constraint-source}, a variety of constraints can be expressed in the application code.

\noindent{\textbf{Database constraints.}} Developer can also declare constraints in the database, such as primary and foreign keys, uniqueness, 
value nullness, and string length constraints.

As shown in previous work~\cite{yang:icse20}, the first category contains very few constraints, while the latter two categories cover more than 
99\% of all constraints. In this paper, we focus on the second category, application constraints, and describe how they can be generalized into 
several common patterns and discovered. As we will see in~\secref{sec:eval}, \textit{most} of these constraints inferred from the second category 
are actually not declared in the database, hence they are not leveraged by the database during query optimization.

\subsection{ORM frameworks}
ORMs provide a high-level API over a relational database. This allows developers to manipulate persistent data and its schemas using the programming language they are familiar with rather than SQL. When executed, ORMs automatically translate each API call into SQL queries.
All ORMs that we are aware of translate such calls straightforwardly and leave query optimization to the underlying database.
This makes sense as each ORM typically supports multiple databases\footnote{Rails for instance supports SQLite, MySQL, PostgreSQL.~\cite{rails-db}} and is thus difficult for it to include optimizations that are compatible with all databases.
Meanwhile, as shown in \tabref{tab:db-rewrite}, none of the popular open-source or commercial database supports all of the query rewriting optimization types provided by \tool, as many of them are application-specific and they take substantial effort to implement without using \tool's ``enumerate-test-verify'' approach.
While \tool’s functionalities can be implemented in the ORM or the database, we implemented our prototype as an independent component as doing so demonstrates that 
our technique is agnostic to any specific ORM or database implementation.
\section{overview}
\label{sec:overview}
We first give an overview of using \tool on an example abridged from Redmine~\cite{redmine}, a popular 
collaboration web application built using Rails.
Redmine defines a {\code User} class to manage user information and a {\code Project} class to store project details. 
The {\code Member} class keeps track of each user's membership information (for example, one user can be a member of many projects).
Rails stores user, member, and project information in separate database tables, and developers manipulate the stored data by calling Rails' functions.
\lstref{code:constraint-redmine} shows the definition of the {\code Member} class on lines 1-4, and the code to create and save a {\code Member} object on lines 6-7. Line 4 utilizes Rails' built-in validation API and is called whenever the object is saved to the database, as shown in line 7. Rails executes the validation on Line 4 by executing a query to determine if a user with the same project already exists in the member table and raises an error if so.
Here, the validation function 
implicitly defines a data constraint that {\em given a project, the users belonging to the project are unique}. 
Yet, it is only defined in the application code but not specified as part of the database schema, as developers can write arbitrary code in the validation function and not all of them can be easily translated to SQL constraints.

\begin{lstlisting}[language=ruby, caption={Redmine code with an implicit data constraint.}, label={code:constraint-redmine}]
# Member Class definition
class Member
    belongs_to :user, :project
    validates_uniqueness_of :user_id, :scope => :project_id
# Create a Member object and save it to the database
member = Member.new(user_id=1, project_id=2)
member.save
\end{lstlisting}

Once the uniqueness constraint is discovered,
\tool leverage it to improve application performance. 
For example, \lstref{code:query-redmine} shows a Redmine query that selects all the active users working on a given project. 
This query has a {\code DISTINCT} keyword to filter out duplicate users.
But, given the constraint mentioned, users working on the same project are guaranteed to be unique, hence there is no need to run duplicate elimination.
With 100K records in the users tables,
this query can be accelerated by 1.65$\times$ by removing the {\code DISTINCT} keyword.
However, we are unaware of any mainstream query optimizer that would perform such optimization as the optimizer is unaware of such constraints.
As we will show, many similar constraints are ``hidden'' in applications that manipulate persistent data, and we are unaware of any existing tool that can discover them. 
Moreover, as we will show in~\secref{sec:eval}, even if we install such ``hidden'' constraints into the database, the optimizer still fails to leverage them to optimize queries
as traditional DBMS uses heuristics to rewrite queries, and they cannot cover all possible optimizations, such as the one shown here. 
\begin{lstlisting}[language=SQL, caption={Query issued by Redmine.}, label={code:query-redmine}]
SELECT DISTINCT users.* FROM users
INNER JOIN members ON members.user_id = users.id 
WHERE users.status = 'active' AND (members.project_id = 2)
\end{lstlisting}

\begin{lstlisting}[language=SQL, caption={Each log record consists of a query template and its parameter values. Here \$1 and \$2 are the query parameters with values ``{\tt User}'' and  ``{\tt AnonymousUser}'' respectively.}, label={code:example-log}]
 SELECT COUNT(*) FROM "users" WHERE "users"."type" IN ($1, $2)  [["User"], ["AnonymousUser"]]
\end{lstlisting}

\begin{figure}
\setlength{\belowcaptionskip}{-10pt}
  \centering
      \includegraphics[scale=0.29]{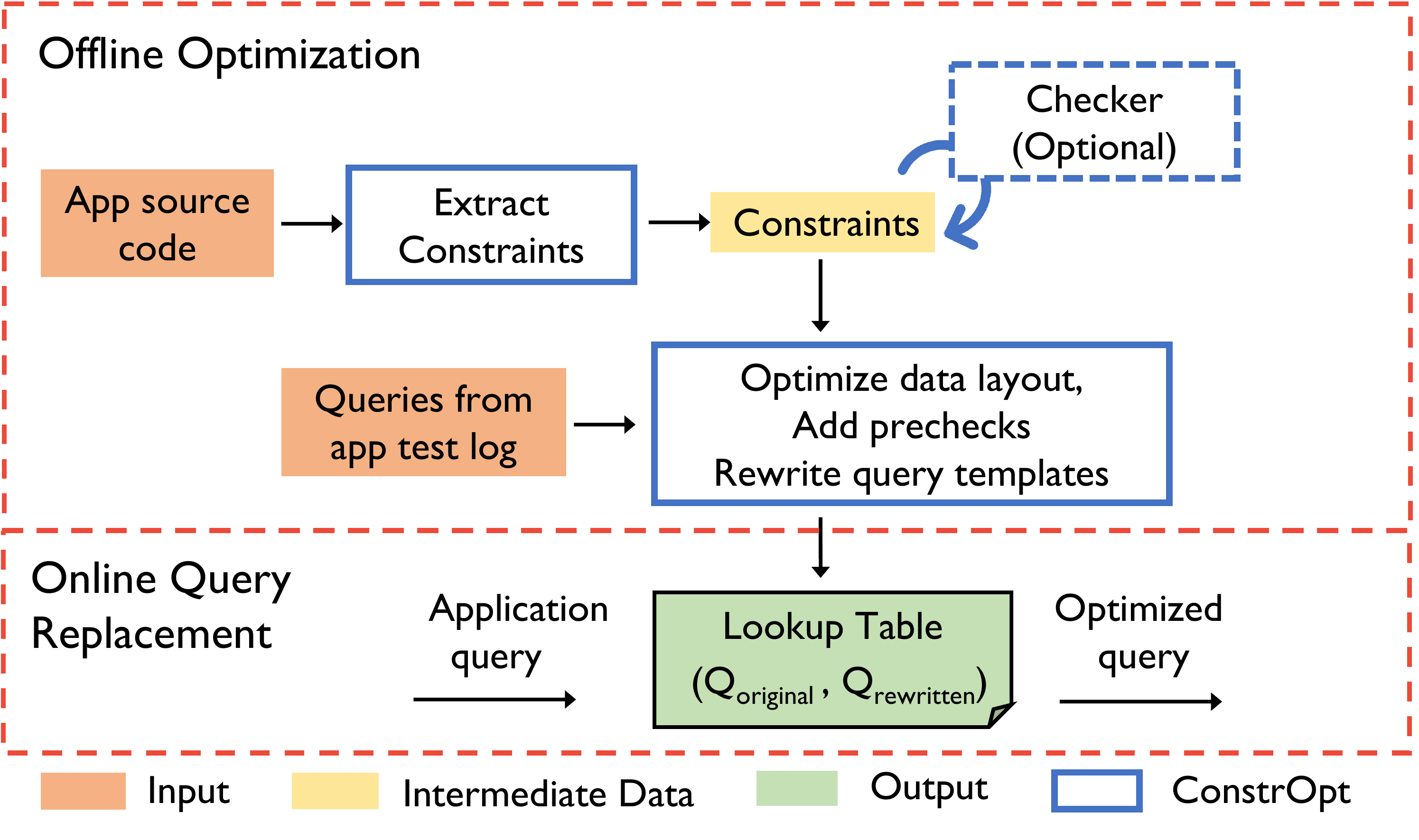}
      \label{fig:arch}
\caption{\tool Architecture.}
\label{fig:arch}
\end{figure}

\tool is built to bridge the gap between application and database. 
It analyzes application source code to automatically extract data constraints.
Moreover, \tool automates the challenging process of leveraging constraints to optimize queries. Even the most sophisticated commercial and open source databases, as demonstrated in~\tabref{tab:db-rewrite}, do not support all of the rewrite types offered by \tool.
\tool optimizes queries using the two components as shown in \figref{fig:arch}.
\begin{itemize} [leftmargin=*,noitemsep,topsep=0em]

\item \textbf{Offline Optimization.} \tool first statically analyzes the application source code to extract constraints.
It detects constraints by matching code patterns defined in~\secref{sec:constraint-source}. The extracted constraints are valid by construction, but \tool also generates a checker program for constraint validation against the stored 
data in case the application violates our assumptions stated in~\secref{sec:assumptions}.
\tool then extracts the queries that the application might issue by analyzing the logs generated by running application tests. \lstref{code:example-log} shows an example log record with two parts: query template and parameters. \tool extracts the query templates and uses extracted constraints to optimize query performance by:
\begin{itemize}
    \item {\em Optimize data layout (\secref{sec:changecodeschema}).} If a string column has a limited set of possible values, i.e., there is an inclusion constraint on the column, \tool changes the data type from string to {\tt enum} to save the storage and speed up queries.\footnote{{In cases where modifying the database schema directly is undesirable (e.g., breaking other applications that access the same database), \tool can also generate a SQL DDL script for the database administrator to determine when to apply the modifications.}}
    \item {\em Add prechecks on query parameters (\secref{sec:changecodeschema}).} \tool utilizes format and length constraints to optimize application logic by adding prechecks on user inputs to avoid issuing queries if the input violates length or format requirements. \tool takes constraints and application source code offline and emits an optimized version with precheck logic.
    \item {\em Rewriting queries (\secref{sec:queryrewrites}).} \tool rewrites query templates to improve query performance using the extracted constraints. To rewrite the query templates, \tool enumerates feasible rewrites (e.g., remove {\code DISTINCT}) based on extracted constraints. It then filters out slow rewrites based on the estimated cost. \tool then uses test cases and a formal verifier to ensure that the rewritten query templates are semantically equivalent to the original. At the end of this offline step, \tool creates a lookup table comprising of the original and optimized query template pairs, and the table is used to rewrite queries online as the application runs.
\end{itemize}
\item \textbf{Online query replacement.}  
As the application runs, \tool intercepts all queries issued by the application. If the query's template exists in \tool's lookup table, \tool issues the optimized query template with its parameters to the database. Otherwise, the query is issued as-is.
\end{itemize}

We now discuss different types of data constraints inherent in the application code and how \tool detects them. Then we introduce the optimizations with these constraints,
followed by an evaluation using real-world database-backed ORM applications. 

\section{Detecting Constraints}
\seclabel{sec:constraint-source}
\begin{table*}
\vspace{-0.3in}
\caption{Rails data validation patterns. In Django~\cite{django-validation} and Hibernate~\cite{hibernate-validation}, all patterns are implemented similarly.}
\label{tab:constraint-pattern}
\centering
\footnotesize
\resizebox{\linewidth}{!}{
    \begin{tabular}{m{10cm} | m{2cm}| m{9cm}}
        \toprule
        \multicolumn{3}{c}{\textbf{Built-in Validation}} \\
        \midrule
            \textbf{Code Pattern} & \textbf{Category} & \textbf{Constraint Description} \\
        \hline
            {\code validates\_inclusion\_of:}\textit{field}, in \textit{value\_list} & Inclusion & \textit{field} takes value from \textit{value\_list}\\
        \hline
            {\code validates\_presence\_of:}\textit{field} & Presence & \textit{field} is not {\code NULL} \\
        \hline
            {\code validates\_uniqueness\_of:} \textit{field}, {\code scope:} \textit{scope\_field} | \textit{scope\_field\_list} & Uniqueness & \textit{(field,scope\_field)} | \textit{[field]} + \textit{scope\_field\_list}  is unique \\
        \hline
            {\code validates\_length\_of | validates\_size\_of:} \textit{field}, \newline
            {\code minimum =>} \textit{value}, {\code maximum =>} \textit{value}, \newline
            {\code in | within} \textit{value\_range} & Length & \textit{field} has type string, and its length is within the given range \\
        \hline
            {\code validates\_format\_of:} \textit{field}, {\code :with =>} \textit{regex} & Format & \textit{field} matches the format specified by the \textit{regex} \\
        \hline
            {\code validates\_numericality\_of:} \textit{field}, {\code greater\_than:} \newline
            \textit{value}, {\code greater\_than\_or\_equal\_to:} \textit{value}, {\code equal\_to:} \newline
            \textit{value}, {\code less\_than:} \textit{value}, {\code less\_than\_or\_equal\_to:} \textit{value} & Numerical & 
            \textit{field}'s numerical value matches the condition specified by the comparison keywords\\
        \toprule
        \multicolumn{3}{c}{\textbf{Custom Validation}} \\
        \midrule
        \texttt{{if} cond \newline
            \hspace*{4mm} errors.add(\textit{error message})} \newline
        where \newline
            \code{
            \hspace*{4mm} cond := expr | op(cond, expr) \newline
            \hspace*{4mm} a := constant | field \newline
            \hspace*{4mm} binop := > | >= | < | <= | == \newline
            \hspace*{4mm} api\_call := length | size | nil? | empty? | blank? \newline
            \hspace*{14mm} | none? | any? | exists? | to\_s | to\_i | to\_f \newline
            \hspace*{4mm} expr := binop(a, a) | api\_call(a) \newline
            \hspace*{4mm} op := || | \&\&} & 
            Custom & \texttt{cond} will produce error, so constraints are generated by taking the negation of \texttt{cond}. \newline
            \tool compares \texttt{field} against the database schema to ensure it is a column persistent in the database and \tool makes sure \texttt{cond} uses at lease one database column. \\
        \bottomrule
    \end{tabular}
}
\end{table*}

\tool extracts both application constraints and database constraints automatically from the source code. Application constraints are embedded semantically in the application code when developers define the model class. Database constraints are specified explicitly in the migration files~\cite{migration}, which are used to alter database schema over time. Constraints defined in migration files will later be installed into the database as database constraints. 

Because of the flexibility, convenience of use, and capacity to manage errors, many constraints are written in the application rather than the database, as indicated in~\secref{sec:eval-constraint-extract}.
Defining constraints in the application code is more flexible as the constraint type is not limited, and developers can write complex logic to express constraints. Meanwhile, constraints not supported by the database must be expressed as user-defined functions (UDFs), which are typically written in SQL and are tedious to use for complex logic. As shown in ~\secref{sec:data-validation}, ORM frameworks also have a number of simple built-in APIs to express common constraints. Finally, developers can associate meaningful error messages when constraints are violated. Whereas the database only throws low-level errors that are rarely caught by the developers~\cite{yang:icse20}. Hence, once triggered, the web user’s session will most likely crash, with all the filled-in contents lost with a cryptic SQL error.

\tool uses both the database constraints and application constraints to improve the query performance.
\tool works by parsing the application source code, building an abstract syntax tree (AST) for each file, and pattern-matching on the AST nodes. 
To simplify the code structure, ConstrOpt extracts data validation, class relationship, and field definition constraints in two separate passes while scanning the AST. Each AST node, which stands for a single source code token, is only visited once each pass. \tool will continue to visit its children if the current node contains any of the patterns listed in Table~\ref{tab:constraint-pattern} and Table~\ref{tab:constraint-pattern-class-relationship}. If there are $n$ tokens in the application source code, the complexity of pattern matching is therefore at most $\mathcal{O}(n)$.

Table~\ref{tab:constraint-pattern} and Table~\ref{tab:constraint-pattern-class-relationship} list all the patterns leveraged by \tool to extract constraints of different types as shown below.
\begin{itemize}[leftmargin=*,noitemsep,topsep=0em]
\item {\bf Inclusion}: the field value is restricted to a limited set.
\item {\bf Presence}: the field value cannot be null. This is the same as SQL {\code NOT NULL} constraint but is implicitly defined in the application.
\item {\bf Length}: the length of a string field should be in a certain range.
\item {\bf Uniqueness}: same as the SQL uniqueness constraint, but is only defined in the application.
\item {\bf Format}: the value of a string field must match a regular expression, which is specified in the application code.
\item {\bf Numerical}: the value of a numerical field must lie in the range specified in the application code.
\item {\bf Foreign key}: same as the SQL foreign key constraint, where the field points to the primary key of the referenced table.
\end{itemize}

For each constraint, we first describe how it conceptually arises from application code, followed by an example, and then the general code pattern that \tool uses for extraction.

\vspace{-0.1in}
\subsection{Data validation}
\seclabel{sec:data-validation}
Data validation is the most important method for extracting the application constraints. 
It's an important feature of ORM-based apps since it ensures that only valid data is kept in the database.
Similar to SQL triggers, most ORMs provide callback mechanisms (e.g., validation functions provided by Rails \cite{rails-validation}, Django validators\cite{django-validation}, Hibernate validators \cite{hibernate-validation}, etc.) where a function is triggered automatically every time before data is saved to the database.
The callback can be one of the ORM's built-in functions that capture common attributes to validate, or one of the user's own validation procedures. 
Inside the callback, a built-in or customized property is checked on the data to be saved, and the function returns an error without saving the data if the check fails. Consequently, such checks lead to constraints that must be satisfied by all stored data.

As an example, \lstref{code:validation} shows two types of validation callbacks from OpenProject~\cite{openproject}.
First, in line 3, a Rails built-in validation function, {\code validate\_format\_of}, is used to
check if the specified field {\code email} satisfies the regular expression in Line 4. 
Then, in line 7, a custom validation function {\code validate\_name}, defined
in lines 8--9, is registered to ensure the length of the {\code name} field 
does not exceed 30 characters, implying a data constraint that {\code length(name)$\leq$30}.

\begin{lstlisting}[language=ruby, caption={Validation function excerpt from OpenProject.}, label={code:validation}, escapeinside={*}{*}]
class User < ApplicationRecord
  # built-in validation
  validates_format_of :email, :with => 
  "/\A([^@\s]+)@((?:[-a-z0-9]+\.)+[a-z]{2,})\Z/i"
  # custom validation
  validate :validate_name
  def validate_name
    if length(name) > 30
      errors.add(:name, "is too long (maximum is 30 characters)")
\end{lstlisting}
\vspace{-0.1in}

To extract data constraints from validation functions, \tool identifies all uses of built-in validations and custom validations registered through {\code validate}. 
As each built-in validation checks for a particular data property, such as {\code validates\_length\_of} checks for length of the passed in field, we define a constraint code template for each built-in validation as shown in \tabref{tab:constraint-pattern}, and use the template to generate a constraint by analyzing the parameter(s) passed to the validation function. 

To handle custom validations, \tool walks through the parsed AST of the function body to identify the branch condition leading to an {\code error} statement. It then matches the branch condition with pre-defined patterns as shown in~\tabref{tab:constraint-pattern},
and uses the negation of the condition as a constraint if the pattern-match succeeds. For instance, in \lstref{code:validation}, \tool identifies that the condition {\code name.length $>$ 30} leads to an error statement shown in Line 10. The branch condition is matched as {\code api\_call(field) > constant} from the grammar.
It then derives a constraint on the {\code User} table as the negation of the condition, i.e., {\code{!(name.length $>$ 30)}}.

\begin{table*}
\vspace{-0.1in}
\caption{Constraint patterns and summary for class relationships and field definition. Django and Hibernate have similar patterns in addition to Rails.}
\label{tab:constraint-pattern-class-relationship}
\centering
\footnotesize
\resizebox{\linewidth}{!}{
    \begin{tabular}{m{9cm} | m{1.5cm}| m{8cm}}
        \toprule
        \multicolumn{3}{c}{\textbf{Class Relationships and Field Definition}} \\
        \midrule
            \textbf{Pattern} & \textbf{Category} & \textbf{Constraint Description} \\
        \hline
            {\code class} \textit{TableClass} {\code < class} \textit{Ci} \textit{(i=1...n)}

           & Inclusion & \textit{C1, C2, ..., Cn} objects will be stored in a single table, where the \textit{type} column is used to differentate records of different classes and takes the value of \textit{Ci(i=1, 2...n)} \\
        \hline
            \textit{TableCalss} {\code belongs\_to} \textit{polymorphic\_key}{\code , polymorphic: true} \newline
                    \textit{Ci} {\code has\_one} \textit{TableCalss}{\code, as} \textit{polymorphic\_key} \textit{(i=1...n)}
                    & Inclusion & 
                    \textit{type} column of \textit{TableClass} can only take value of \textit{C1, C2...Cn} \\
        \hline
            \textit{ClassA} {\code has\_one} \textit{ClassB}{\code, (-> \{where} \textit{field} \textit{op} \textit{value} {\code\}) } & Uniqueness & \textit{ClassB} has a column named \textit{ClassA\_id}, which points to ids of \textit{ClassA}. A uniqueness constraint is introduced here where among all tuples that match the where predicate, the value of \textit{ClassA\_id} field is unique.\\
        \hline
            {\code state\_machine:} \textit{field}{\code, initial:} \textit{initial\_state},
            {\code event:} \textit{event\_name},
            {\code transition from} \textit{event\_start\_states} {\code to} \textit{event\_end\_states}
            & Inclusion & \textit{field} takes value from [\textit{initial state}] + \textit{event\_start\_states} + \textit{event\_end\_states}\\
        \bottomrule
    \end{tabular}
}
\end{table*}

\subsection{Class relationships}
ORM frameworks support complex relationships between classes such as class hierarchies, polymorphic one-to-many, etc. The framework maintains those relationships, which can be found as constraints on persistent data. The specific patterns are listed in ~\tabref{tab:constraint-pattern-class-relationship}.

\subsubsection{Type hierarchy.}
As discussed in \secref{associations}, ORMs use different mechanisms to support inheritance in relational databases, such as one table per entire class hierarchy, or one table per class.
By default, Rails~\cite{rails-inheritance} (and similarly Django~\cite{django-inheritance}) uses the table per hierarchy mechanism where an additional string field to store the classname is used to differentiate different class instances. 
This field is called {\code type} by default, or can be explicitly defined by the user in the {\code inheritance\_column}.
Similar to the example in \lstref{code:inheritance}, where both \texttt{Firm} and \texttt{Client} inherit from \texttt{Company}. Rails keeps a single {\code companies} table to store instances of both {\code Firm} and {\code Client}, with the {\code type} field of this table indicating which class the record belongs to. This mechanism introduces an inclusion constraint, where the value of the {\code type} field can only be {\code Firm} or {\code Client}. 

\begin{lstlisting}[language=ruby, caption={Example of type inheritance, with bodies of the class definitions omitted due to space.}, label={code:inheritance}]
class Company < ActiveRecord::Base;
class Firm < Company;
class Client < Company;
\end{lstlisting}


To extract class inheritance constraints, \tool maps each class to its inherited classes by searching for class definitions in the code.
\tool then generates inclusion constraints for each entry in the mapping where the key is the table and column name, and the value is the value range of the inclusion constraint. For instance, in the case shown in \lstref{code:inheritance}, \tool will detect that {\code Company.type} can only have values {\code `Firm'} or {\code `Client'}.

\subsubsection{Polymorphic definitions}
Similar to {\code type} fields in class inheritance, ORM frameworks allow developers to define a field that refers to the primary key of multiple tables, and uses an extra string field to identify which table the record belongs to.
An example is shown in~\lstref{code:polymorphic}: each {\code Organization} and {\code User} object contains a single {\code Address}, and hence an {\code Address} object can belong to either an {\code Organization} or a {\code User}. This polymorphic relationship is declared in line 2, where each {\code Address} is set to belong to an {\code addressable} interface, and the {\code Organization} and {\code User} classes are declared as {\code addressable} in lines 5 and 8.

Internally, since all {\code Address} records are stored in a single {\code address} table, Rails adds an integer field {\code addressable\_id} which refers to a primary key in either the {\code organizations} or {\code users} table, and a string field {\code addressable\_type} to store the type name of the object that an {\code address} belongs to. 
This mechanism allows for inclusion constraints where {\code addressable\_type} can only take the value of {\code `Organization'} or {\code  `User'}. Similar to the type hierarchy, \tool automatically analyzes and infers such constraints. 

\tool also extracts polymorphic definitions by building a mapping from the polymorphic class name and the interface name (e.g., {\code (Address, addressable)}) to a list of classes that use the interface (e.g., {\code [`Organization', `User']})

\begin{lstlisting}[language=ruby, caption={Example of a polymorphic interface definition.}, label={code:polymorphic}]
class Address < ActiveRecord::base
  belongs_to :addressable, polymorphic: true

class Organization < ActiveRecord::base
  has_one :addresses, as: :addressable

class User < ActiveRecord::base
  has_one :addresses, as: :addressable
\end{lstlisting}

\subsubsection{has\_one association.} 
ORMs provide {\code has\_one} associations to express that exactly one other class has a reference to a class object. \lstref{code:relation-simple} shows an example, where line 2 declares a {\code WikiPage} object belonging to a {\code Project} object, while each {\code Project} has only one {\code WikiPage} (line 5). With the {\code belongs\_to} association, a foreign key field {\code project\_id} is added to the {\code wikipages} table. 
This implies that the constraint {\code project\_id} is unique across the {\code wikipage} table. 

\begin{lstlisting}[language=ruby, caption={Example of data association declaration.}, label={code:relation-simple}]
class WikiPage < ActiveRecord::base
    belongs_to project, class_name: 'Project'

class Project < ActiveRecord::base
    has_one wikipage, class_name: 'WikiPage'
\end{lstlisting}


\tool extracts such associations by matching the {\code has\_one} keyword for each class, and records its {\code class\_name} and the associated class.
A unique constraint is then generated on the {\code class\_name\_id} field of the table identified by the associated class name.

\subsection{Field definition with state machines}
Instead of issuing an {\code UPDATE} query, applications often use libraries to change the value of a field. For instance, the {\code state\_machine} library is commonly used to define how the value of a field can be changed~\cite{state-machine1}. An example is shown in \lstref{code:statemachine} from Spree~\cite{spree}, where changing the {\code state} field is done by a {\code state\_machine}: as shown in line 2, the value of {\code state} can be changed from {\code `checking'}, {\code `pending'}, {\code `complete'} to {\code `processing'} only when the \newline
{\code start\_processing} event is triggered. 
Using {\code state\_machine} to update {\code state} implies a constraint where the value of {\code state} can only be one of the string literals defined within the state machine. 

\begin{lstlisting}[language=ruby, caption={Example of field definition using the state machine library from Spree.}, label={code:statemachine}, linewidth={9cm}]
class Payment
  state_machine : state, initial: 'checkout' do
    event :start_processing do
      transition from:['checkout','pending','complete'], 
            to: 'processing'
    event :failure do
      transition from:['pending', 'processing'], to:'failed'
\end{lstlisting}

\tool extracts state machine constraints by first identifying {\code state\_machine} library calls. 
It 
goes through each of the event functions and extracts all parameters in the {\code from:} and {\code to:} expressions to obtain all the potential state values. If all the values are string literals, \tool will generate a corresponding inclusion constraint. In using state machines, the state variable can only be changed when the specified event takes place 
\cite{statemachinedoc}. Therefore, all possible states must appear within the same {\code state\_machine} construct, as shown in \lstref{code:statemachine}.

\vspace{-0.1in}
\section{Validity of our approach}

We first describe the requirements for applications to use \tool. Next, we list the assumptions \tool makes, and how \tool can handle different situations pertaining to application changes.
\vspace{-0.1in}
\subsection{Requirements and Assumptions}
\label{sec:assumptions}

\tool extracts constraints through pattern matching on the validation APIs defined by the ORM. While \tool currently focuses on the Rails ORM, the validation APIs defined by other ORMs such as Hibernate and Django are essentially identical Rails'. Therefore, a large number of ORM applications should readily benefit from \tool.

Moreover, to guarantee the correctness of extracted constraints, we make the following assumptions about ORM applications:
\begin{itemize}[leftmargin=*,noitemsep,topsep=0em]
    \item Data validation, as discussed in \secref{sec:constraint-source}, is not bypassed. Even though developers can skip validation when saving an object by setting the validate parameter to false (e.g., {\code obj.save(validate: false}), such behavior violates the design principle of validation and is highly discouraged~\cite{skipvalidation}. We scan the code of all 14 applications used in evaluation and only find two cases where the developer skips the validation. Those two cases update only fields that are irrelevant to validation, and therefore do not affect the validity of the extracted constraints.
    \item The application does not use any non-analyzable third-party library that breaks existing constraints.
    We assume that the application code is analyzable, as  
    we cannot guarantee constraint validity if a third-party library whose source code is not available modifies the fields involved in the constraints. However, this case is rare and we did not find any in the evaluated applications.
\end{itemize}

Given these assumptions, \tool will generate valid constraints. While we focus on the database being accessed by a single application, if there are multiple applications accessing the same database, 
\tool combines the constraints such that they hold for {\em all} applications to ensure the validity of any resulting query rewrites as follows:
\begin{itemize}[leftmargin=*,noitemsep,topsep=0em]
\item \textbf{Inclusion.} If there are more than one inclusion constraint on the same field, e.g., one app defines field \texttt{f} to take the value from \texttt{(A,B,C)}, and another app requires \texttt{f} to be chosen from \texttt{(B,C,D)}, \tool generates a new inclusion constraint on {\tt f} stating that its value is the intersection of the two, i.e., {\tt (B,C)}.
\item \textbf{Format.} If a field has several format constraints, \tool combines them by stating that the field must adhere to the conjunction of all format constraints.
\item \textbf{Length/Numerical.} If there is more than one length/numerical constraint on the same field, \tool generates a new constraint that sets the length/value limit of the field to be the intersection of the length/value ranges. For instance, if one constraint states that field {\tt f} $\in$ \{1,10\} and another states that {\tt f} $\in$ \{0,5\}, then \tool intersects them and states that {\tt f} $\in$ \{1,5\}.
\item \textbf{Presence/Uniqueness/Foreign Key.} \tool only keeps the presence/uniqueness/foreign key constraints on a field if the same constraint is extracted from all applications.
\end{itemize}

\vspace{-0.1in}
\subsection{Code upgrade}
Developers can run \tool to re-detect constraints when application code is updated. As shown in~\secref{sec:eval-opt-detect}, re-extracting constraints is fast and efficient. Moreover, incremental constraint detection can be performed by scanning only the modified files to further reduce the time needed to extract constraints. Note that code changes might lead to adding new or altering previously extracted constraints that are incompatible with old data. As shown in previous work~\cite{yang:icse20}, the mismatch between constraints and data is problematic and developers should ensure that data integrity is preserved.
Multiple methods have been proposed to fix this. For example, migration files can be used to ensure that all data in the database satisfy the newly inferred constraints.
\tool's checker program described in~\secref{sec:constraints-checker} can also help developers identify such data integrity problems and ensure the validity of extracted constraints.

\vspace{-0.05in}
\subsection{External changes and constraints checker}
\label{sec:constraints-checker}
In cases where the DB contents are not only modified by the application, such as when a DB administrator manually changes the contents of the database via a command line interface, the constraints detected by \tool may become invalid. Although such case is discouraged and
rare, \tool comes with a checker that examines if constraints still hold after a third party modifies the database. The checker can be used to validate the extracted constraints under the two assumptions described in \secref{sec:assumptions} as well. 
For every extracted constraint, \tool generates a Ruby script that can be used to scan all the data from the database and check whether the extracted constraints are valid.
The checker script can be run against a concrete database instance and remove any \tool-extracted constraints that are no longer valid.
In general, we expect such cases to be rare and leave the decision of when to run the checker script to the developer.

\vspace{-0.1in}
\section{Query Analysis and Optimization}
We now discuss how \tool rewrites queries with extracted constraints. 
We summarize the constraints used in different optimization 
in~\tabref{tab:constraint-rewrite},
and discuss how \tool leverages them to optimize queries by changing the source code or database schema and rewriting queries with our enumerate-test-verify approach. 

\vspace{-0.1in}
\subsection{Query extraction}
\seclabel{sec:queryextraction}
\tool extracts SQL queries that can be issued by the application by running its provided test cases and analyzing the log file that records all SQL queries executed. 
\tool replaces the issued query template (as discussed in~\secref{sec:overview}) with the optimized version if it exists in the lookup table. 
As the application runs, queries are issued against  different query templates after pairing them with parameters.
Therefore, ConstrOpt can still apply the optimization as long as the query template exists in its lookup table.
Furthermore, as tests are carefully written with a high level of code coverage (over 92\% for the applications used in our evaluation), \tool should have analyzed most templates that the application can possibly issue when deployed.

\vspace{-0.1in}
\subsection{Optimizing code logic and physical design}
\seclabel{sec:changecodeschema}

\tool leverages the extracted constraints to optimize performance by rewriting application code and changing the datatype of underlying storage. We describe both below.

\noindent\textbf{Adding prechecks to avoid issuing queries.}
A query can be completely removed if it returns empty results. For example, for a query containing a predicate that compares a field with user input, we can add a precheck 
in the application code 
to issue the query only if the input matches the field's associated constraints.
\lstref{code:devto-precheck} shows an example from Dev.to~\cite{devto}, where \tool adds precheck on the string field {\code username}. Here \tool extracts a constraint that all characters in {\code username} must be digits, letters, or underscores, as expressed by matching the regular expression in line 1. For the query that retrieves users given  username (line 2)
, we add a precheck to the parameter (line 4) that avoids issuing the query altogether if the parameter contains invalid characters and hence is impossible to 
match a username in the database.

\begin{lstlisting}[language=ruby, caption={Adding parameter check example from Dev.to~\cite{devto}. {\color{green}+} indicates code added by \tool.}, label={code:devto-precheck}]
@\color{green}+@ if param[:username].match(/\A[a-zA-Z0-9_]+\z/)
     user = User.where(username: param[:username])
@\color{green}+@ else
@\color{green}+@    user = nil
\end{lstlisting}
\vspace{-0.1in}


To implement this, \tool checks each query $Q$ for the presence of the predicate {\code TableName.field={\it param}}. Using the data flow graph, \tool then traces if {\it param} is computed from the user input and whether there is a \tool-extracted constraint associated with {\code field}. If $Q$ satisfies these criteria, \tool adds the precheck as shown in~\lstref{code:devto-precheck}. 


\noindent \textbf{Altering database schema.}
If a string field has an inclusion constraint (i.e., its value can only come from a limited set of literals), we change its physical design 
and replace the field with an enumeration type. Since {\tt enum} comparison is faster than string comparison,  changing the storage can reduce the time to process predicates on that field. Additionally, it improves space efficiency, as the {\tt enum} type is stored using only four bytes in the database~\cite{enum-psql, enum-mysql}.

To implement this, \tool checks all extracted inclusion constraints and changes the type of field {\code f} to enumeration type if there is an inclusion constraint on the field. 
ConstrOpt creates an {\tt enum} datatype based on extracted inclusion constraints. It then generates {\tt ALTER TABLE} statements that change {\tt f}'s type from  {\tt varchar} to {\tt enum} based on the value list of the inclusion constraint. \tool can execute those statements directly or return them to the DB administrators to determine when to apply them (e.g., when there are no updates to the affected columns). Queries involving field {\code f} will remain unchanged as the database will perform type conversion when running the query.

Notice that the above mentioned optimizations can only be performed by \tool: the database is unaware if a query parameter is derived from user inputs and is unable to add prechecks on input columns. Moreover, the database is also unaware of inclusion constraints and thus cannot alter its schema automatically. Using \tool, however, a large number of queries can benefit from the two optimizations as we will discuss in~\secref{sec:eval-opt-detect}.

\vspace{-0.1in}
\subsection{Rewriting Queries}
\seclabel{sec:queryrewrites}
The previous two optimizations do not need modifying the query, however many queries do require exploiting extracted constraints to change the query to a semantically equivalent but more efficient one.
In~\tabref{tab:constraint-rewrite}, we outline the taxonomy of using constraints for query optimizations in \tool. 

We first install the \tool-extracted constraints into the database and see if the database can leverage them to optimize queries. However, as shown in~\secref{sec:eval-perf}, only a few queries can benefit even after the constraints are in place. Most major database management systems, as shown in~\tabref{tab:db-rewrite}, are unable to leverage \tool-extracted constraints to optimize queries.
The reason is that existing DBMS 
relies on heuristics to exploit constraints and rewrite queries. Each type of rewrite necessitates different rules, and some of which are rather complex.
For instance, to remove the {\code DISTINCT} keyword in a query after identifying a uniqueness constraint, the optimizer must track the uniqueness
of all the columns used in every operator, since not all query operators preserve the uniqueness of the input (e.g., projection)~\cite{pg-remove-distinct}.
Implementing such rules requires substantial effort, yet it is unclear how generally applicable they are. For instance, in the Redmine only 399 out of 2283 queries use {\code DISTINCT}, among which only 67 can be removed.

\begin{table}[t]
\centering
\caption{Rewrite types supported by mainstream DBMS.}
\label{tab:db-rewrite}
\resizebox{0.96\linewidth}{!}{
\begin{tabular}{c|c|c|c|c|c } 
\toprule
  DBMS & \shortstack{Remove\\Distinct} &  \shortstack{Add Limit\\One} &  \shortstack{Predicate Elimination\\Introduction} & \shortstack{Join Elimination\\Introduction} & \shortstack{Detect\\Empty Set} \\
\midrule
PostgresSQL 10.7  & $\times$    & $\checkmark$ & $\times$     & $\times$     & $\times$ \\
MySQL 8.0.2       & $\times$     & $\checkmark$ & $\times$     & $\times$     & $\times$ \\
SQL Server 2019   & $\times$     & $\times$     & $\times$     & $\checkmark$ & $\times$ \\
DB2 10.5 Kepler & $\times$     & $\times$     & $\checkmark$ & $\checkmark$ & $\times$ \\
\tool             & $\checkmark$ & $\checkmark$ & $\checkmark$ & $\checkmark$ & $\checkmark$ \\
\bottomrule
\end{tabular}}
\end{table}

\begin{algorithm}
\caption{Rewrite generation, test, and verification algorithm }
\label{alg:rewrite-test-verify}
\begin{algorithmic}[1]
\Require Original query template $Q$, and all constraints $C$.
\Ensure Equivalent rewrite $R$ with the minimum cost.

\State \textbf{\em // Step 1: enumerating potential rewrites.}
\State $fields$ = get\_used\_fields($Q$)
\State $CQ$ = get\_constraints\_on\_fields($fields$, $C$)
\State $rewrite\_types$ = get\_constraint\_rewrites(CQ)
\State $R_{enumerate} = \{Q\}$
\For{$rt \in rewrite\_types$}
    \State $R_t = \{\}$ \textit{// rewrites after applying $rt$}
    \For{$candidate \in R_{enumerate}$}
        \State $R_t$ += apply\_transformation($rt$, $candidate$)
    \EndFor
    \State $R_{enumerate}$ += $R_t$
\EndFor

\State \textbf{\em // Step 2: Cost estimation and testing rewrites.}
\State $R_{cost} = \{\}$
\For{$candidate \in R_{enumerate}$}
    \If {cost(instantiate($candidate$)) < cost(instantiate($Q$))}
    \State $R_{cost}$.add($candidate$)
    \EndIf
\EndFor
    
\State $R_{test} = \{\}$
\For{$candidate \in R_{cost}$}
    \If {test\_eq\_on\_test(instantiate($candidate$), instantiate($Q$))}
        \State $R_{test}$.add($candidate$)
    \EndIf
\EndFor
\State {\em // Sort rewrites based on cost in ascending order}
\State $R_{test\_sort}$ = $R_{test}$.sort(key = cost, asc=True)

\State \textbf{\em // Step 3: formally verifying rewrite equivalence.}
\For{$candidate \in R_{test\_sort}$}
\If {verify\_eq($candidate$, $Q$)}
    \State Return $candidate$ {\em // Early stop}
\EndIf
\EndFor
\\
\Return NULL {\em // fail to find an optimized rewrite}
\end{algorithmic}

\end{algorithm}

\tool instead uses an enumerate-test-verify approach as shown in Algorithm~\ref{alg:rewrite-test-verify} to automate the rewrite process. First, \tool enumerates all possible transformations of a query template given the rewrites shown in ~\tabref{tab:constraint-rewrite}.
For each query template, we restrict the maximum number of enumerated rewritten templates to 200. As described in ~\secref{sec:eval-opt-detect}, only a few query templates exceed this limit for the six applications that are evaluated.
To reduce the number of potential rewrites, only query templates containing fields with constraints are enumerated. \tool then puts the parameter values into the rewritten templates and uses the query optimizer to estimate the cost of each instantiated rewrite.
Rewrites with a lower cost than the original are then checked for semantic equivalence. As formal verification can be costly, \tool instead generates a test database to execute each instantiated rewrite and original query. \tool filters away those cases where the instantiated rewrite returns different results from the original.
\tool then sorts the remaining rewrites 
based on their estimated costs in ascending order and sends them to our formal verifier to check for query equivalence. \tool finally emits the rewritten template once it finds the first equivalent one. We now describe these steps in detail.

\begin{table*}[t]
\centering
\caption{Description of rewrite types, the constraints that trigger the rewrite, and the implementation.}
\footnotesize
\resizebox{\linewidth}{!}{
\begin{tabular}{m{2cm}|m{1.6cm}|m{8cm}|m{10cm}} 
\toprule
{\bf Rewrite Type} & {\bf Constraints} & {\bf Enumeration implementation} & {{\bf When is the enumeration correct and why is it beneficial}}\\
\midrule
Remove DISTINCT~\cite{habimana2015query}
    & Uniqueness
    & Remove the {\code DISTINCT} keyword if there is any uniqueness constraint on any of the used columns in the query.
    & The selection result is known to be unique. It is advantageous because the {\code DISTINCT} operator sorts or aggregates data, both of which are expensive operations. \\
\hline
Add LIMIT One~\cite{yang2018not}
    & Uniqueness       
    & Add {\code LIMIT 1} to the end of the query and any subquery if there is any uniqueness constraint on any used columns in the query.
    & If no more than one record will be selected, add {\code LIMIT 1} to avoid unnecessary
    operations (e.g., scan) after finding one satisfying record. This is useful if there is no index on the unique column, and a sequential scan must be performed to find the matching result. \\
\hline
Predicate\newline Elimination~\cite{cheng1999implementation, rewrite-theory1} 
    & Numerical or Presence  
    & Remove the predicate if there are any numerical constraints or presence constraints on the fields included in the predicate.
    & The predicate is known to be always true given the constraints. This is beneficial because of avoiding unnecessary predicate comparison.\\
\hline
Predicate\newline Introduction~\cite{cheng1999implementation, rewrite-theory1, rewrite-theory4}
    & Numerical
    & Use a solver~\cite{z3} to enumerate all non-redundant formulas derivable from the predicates and numerical constraints, and add them to the predicate if there is a numerical constraint on any of the used columns in the query.
    & The added predicated is guaranteed to be true. A new predicate on an indexed attribute may allow for a more efficient access method. Similarly, a new predicate on a join attribute may reduce the number of join records, thus improving join performance. \\ 
\hline
Join\newline Elimination~\cite{cheng1999implementation, rewrite-theory1, rewrite-theory4}
    & Foreign Key and Presence  
    & Enumerate all possible ways to drop the join table and the join conditions if there is any foreign key constraint on any used columns in the query.
    & A join may be constrained such that its result is known a priori and does not need to be evaluated. For example, queries that join two tables related through a referential integrity constraint. \\
\hline
Detecting the Empty Answer Set~\cite{cheng1999implementation, rewrite-theory1}
    & Numerical or Presence
    & Modify the predicate to {\code False} if there are any numerical constraints or presence constraints on the fields included in the predicate.
    & If the query predicates are inconsistent with the integrity constraints, the query result is always empty, and we can avoid issuing the query completely. \\
\bottomrule
\end{tabular}
}
\label{tab:constraint-rewrite}
\vspace{-0.1in}
\end{table*}

\noindent\textbf{Step 1: Heuristic-guided rewrite.}
We describe the query optimizations with constraints that \tool leverages in~\tabref{tab:constraint-rewrite}. 
As described, each rewrite leverages certain types of constraints. Therefore, \tool enumerates potential rewrites only when the corresponding constraints exist. As shown in Algorithm~\ref{alg:rewrite-test-verify} lines 2--10, \tool extracts all columns used in a query template and checks them against the extracted constraints. For each existing constraint, \tool applies the corresponding potential transformations as shown in~\tabref{tab:constraint-rewrite}. For instance, for the query in \lstref{code:heuristic-rewrite}, \tool first extracts the used columns {\code members.user\_id}, {\code users.id, users.status}, and {\code members.project\_id}.
\tool then determines that 
each {\code users.id} and each pair of 
({\code members.user\_id}, {\code members.project\_id}) is unique. It then applies the remove {\code DISTINCT} and add {\code LIMIT 1} transformations and generates three candidate rewrites as shown in lines 14--16.
\tool's modular design makes it easy to add new types of rewrite rules. Users can simply add a new semantic query rewrite to the search space by providing the utilized constraints and associated enumeration.

The candidate rewrites generated by the enumeration step are not guaranteed to be semantically equivalent to the original or to perform better, and that is the goal of Step 2.

\begin{lstlisting} [language=SQL, caption={Heuristic-guided rewrite for a Redmine query.}, label={code:heuristic-rewrite}, linewidth={9cm}]
-- Original Query
SELECT DISTINCT users.* from users
INNER JOIN members ON members.user_id = users.id
WHERE users.status = $1 AND (members.project_id = $2)

-- Used columns
members.user_id, users.id, users.status, members.project_id

-- Extracted constraints on used columns
Uniqueness: (members.user_id, members.project_id) pair is unique
Uniqueness: users.id is unique

-- Candidate rewritten templates: 
1. SELECT users.* from users INNER JOIN members ON members.user_id = users.id WHERE users.status =  $1 AND (members.project_id = $2)  -- remove DISTINCT
2. SELECT DISTINCT users.* from users INNER JOIN members ON members.user_id = users.id WHERE users.status =  $1 AND (members.project_id = $2) LIMIT 1  -- add LIMIT 1
3. SELECT users.* from users INNER JOIN members ON members.user_id = users.id WHERE users.status =  $1 AND (members.project_id = $2) LIMIT 1  -- remove DISTINCT and add LIMIT 1
\end{lstlisting}

\noindent\textbf{Step 2: Cost estimation and check for rewrite equivalence using test database.}
For each enumerated candidate template, \tool first instantiates it by binding the parameter values as the original query. Next, \tool asks the DB optimizer to estimate its cost, as shown in lines 13--16, and only retains it if it has a potentially lower cost than the original query.
After filtering away the slow rewrites, \tool then attempts to eliminate as many incorrect rewrites as possible to reduce the number of rewrites that need to be verified. To achieve this, \tool generates a synthetic database given the table schema and compares the outputs of the candidate and the original query. The templates of rewrites that produce the same outputs as the original query are sent to step 3 for verifying query equivalence formally.


\noindent\textbf{Step 3: Formal verification of the candidate rewrites.}
As the last step, \tool calls its verifier to check the equivalence of the original query template and rewritten templates for all remaining candidates.
The verifier is based on Cosette's U-semiring semantics~\cite{cosetteVLDB} and implements the U-semiring decision procedure (UDP).
Given a pair of query template to check, UDP translates each SQL query template to an expression of U-semiring, rewrites into sum-product normal form (SPNF, as given in \cite{cosetteVLDB}), and finally attempts to unify both SPNFs.
The UDP approach already models uniqueness, key constraints, and treats aggregations as uninterpreted functions.
To model the {\code LIMIT 1} transformation, we introduce a new axiom for the {\code LIMIT} operator.
\begin{equation*}
    \sum_a R(a) = \Big \lVert \sum_a R(a) \Big \rVert \implies \mathrm{Limit}(1, R) = R.
\end{equation*}
This captures the idea that when there are no more than one rows in $R$, we can turn $\mathrm{Limit}(1, R)$ into $R$.

We model constraints on columns by adding a predicate over columns, which will be injected as additional predicates in the U-semiring expressions.
For example, the numerical constraint of less than $100$ for some column $k$ in table $R$ is handled by the solver with the rewrite
$R(k) = [k < 100] \times R(k)$ after normalization.
The verifier then calls an SMT solver~\cite{z3} to check for logical equivalence between the predicates during the unification of SPNFs.

\noindent{\textbf{Complexity Analysis.}
\begin{table}[]
    \centering
    \caption{Notation used in complexity analysis}
    \vspace{-0.1in}
    \label{tab:notation}
    \resizebox{0.8\linewidth}{!}{
        \begin{tabular}{c|c|c|c}
        \toprule
        Symbol & Description & Symbol & Description \\
        \midrule
        $Q$ & Original query & $n$ 
            & Length of $Q$ \\
        $C$ & Num. of constraints on $Q$ & $R$
            & Num. of rules to apply \\
        $k$ & Num. of rewrites with one rule
            & $f_{AR}$ & Cost of applying one rule \\
        $f_{c}$ & Cost of estimating cost
            & $f_{test}$ & Cost of running tests\\
        $f_{veri}$ & Cost of verifying query equivalence \\
        \bottomrule
    \end{tabular}}
\vspace{-0.2in}
\end{table}
We list the notation used in~\tabref{tab:notation}. The first step in enumerating potential rewrites includes two operations: get the rules to apply and enumerate all possible ways of applying those rules on $Q$. To get the rules, \tool first scans the query $Q$ and extracts all the fields used in $Q$. This operation takes $\mathcal{O}(n)$ time. Next, \tool checks if there are any constraints on the used fields, which takes $\mathcal{O}(n)$ time if we use a hash table to store all constraints. Then \tool extracts the rules based on constraints on the query using the rules from \tabref{tab:constraint-rewrite}, which takes $\mathcal{O}(C)$. Lastly, \tool uses breadth-first search (lines 6-10 in Algorithm~\ref{alg:rewrite-test-verify}) to enumerate all possible ways of applying those rules on $Q$. As there are $\mathcal{O}(k^{R})$ potential rewrites and enumerating each rewrite takes $f_{AR}(n)$ time, the total cost of this step is $\mathcal{O}(2n + C + k^{R} f_{AR}(n))$.}

The second step tries to estimate the execution cost of each candidate, checks the rewrite correctness by running tests against the database, and sorts the rewrites based on cost. There are $\mathcal{O}(k^{R})$ candidate rewrites and the complexity to estimate cost and run tests are $f_c$ and $f_{test}$ respectively. Therefore the complexity of the first two operations are $\mathcal{O}(k^{R} (f_c(n)+f_{test}(n)))$. Next, the sort operation on $\mathcal{O}(k^{R})$ candidate rewrites has complexity $\mathcal{O}(n k^{R}log(k^{R}))$. The total cost of step two is $\mathcal{O}(k^{R} (f_c(n)+f_{test}(n)+n log(k^{R})))$.

The last step verifies the equivalence of each rewrite pair. Since there are $\mathcal{O}(k^{R})$ candidate rewrites and the cost to verify each pair is $f_{veri}$, the total cost of this step is $\mathcal{O}(k^{R} f_{veri}(n))$.

Therefore, the total complexity of Algorithm 1 is the sum of the three steps above.
That is:
\[
\mathcal{O}(2 n + C + k^R(f_{A R}(n) + f_c(n) + f_{test}(n) + n \log(k^R) + f_{veri}(n))).
\]

From the experiments as described in \secref{sec:eval-opt-detect}, we know that $k^R$ is less than 200 for most queries extracted from evaluated applications. Therefore, the complexity of Algorithm 1 is manageable for real-world applications given $f_{A R}$, $f_c(n)$, $f_{test}(n)$, $f_{veri}(n)$ are all small as shown in \secref{sec:eval-opt-detect}.

\noindent{\bf Data Generation.}
\seclabel{sec:data-gen}
We created a data generator to create synthetic data.
Data generation is done in 3 phases. First, \tool creates a graph of Rails models where the vertices represent Rails models, and the edges denote the dependencies between models (e.g., foreign key). \tool then runs the topological sort on the graph to generate the order to populate test data.
\tool then generates a ruby file for each Rails model that inserts data into the DB when executed.
Data is uniformly generated subject to the extracted restrictions. For instance, for strings, we first check if there is a length constraint on the given field. If so, \tool will create strings with random characters and random length that fall within the bounds of the constraint. 
Finally, \tool executes each generated ruby file based on the topological sort order from the Sorting phase. Since we insert data through the Rails API, all generated data will be validated before inserting into the database.
\vspace{-0.2in}
\section{Evaluation}
\seclabel{sec:eval}

\subsection{Experiment setup}

\noindent{\bf Application corpus.}
We select 14 real-world web applications built using the Ruby on Rails framework~\cite{rails} that cover 5 categories: forum, collaboration, 
e-commerce, social network, and map application. We select the most popular, actively maintained applications from each category based on the number of GitHub stars as listed in~\tabref{tab:apps}. All the applications have over 1K stars and have been developed for more than 
four years. 

\begin{table}
\caption{Details of the applications chosen in our study. Fields show the number of average fields across all tables.}
\label{tab:apps}
\scriptsize{
  \vspace{-0.1in}
  \begin{centering}
  \scalebox{1.1}{
  \begin{tabular}{c|ccccc}
  \toprule
  	{\bf Category}              & {\bf Abbr.}  & {\bf Name}      & {\bf Stars}  &  {\bf Tables}  & {\bf Fields} \\
  \midrule
	\multirow{4}{*}{Forum}& Ds     & Discourse\cite{discourse} & 30.8k   & 180    & 9  \\
                          & Dv     & Dev.to\cite{devto}        & 2.1k    & 92     & 13 \\
                          & Lm     & Loomio\cite{loomio}       & 1.9k    & 50     & 19 \\
                          & Lb     & Lobsters\cite{lobsters}   & 3.0k    & 15     & 4  \\
  \hline
  \multirow{3}{*}{Collaboration} & Re & Redmine\cite{redmine}           & 3.7k    & 54     & 8  \\ 
                                 & Gi & Gitlab\cite{gitlab}             & 22.2k   & 337    & 8  \\
                                 & Op & OpenProject\cite{openproject}   & 1.2k    & 114    & 7  \\
  \hline
  \multirow{2}{*}{E-commerce}    & Ro & Ror ecommerce\cite{ror-ecommerce}  & 1.2k    & 65   & 6  \\
                                 & Sp & Spree\cite{spree}                  & 11.4k   & 57   & 11 \\
  \hline
  \multirow{4}{*}{Social \newline Network}  & Da & Diaspora\cite{diaspora} & 12.4k & 50 & 7  \\
                                            & On & Onebody\cite{onebody}   & 1.4k  & 57 & 11 \\
                                            & Ma & Mastodon\cite{mastodon} & 3.6k  & 78 & 6  \\
                                            & Ta & Tracks\cite{tracks}     & 1.0k  & 17 & 6  \\
  \hline
  Map Applications & Os & OpenStreetmap\cite{openstreetmap} & 2.1k & 46 & 7\\
 \bottomrule
\end{tabular}}
\end{centering}
}
\end{table}

\noindent{\bf Evaluation platform.}
We implement \tool using Ruby, Python, and Rust, where the Ruby code analyzes the application code to detect constraints, the Python code enumerates and filters out slow rewrites, and the Rust code formally verifies query equivalence. We run the applications on AWS c5.4xlarge instance with 16 vCPUs and 32GB memory to measure query performance and use PostgresSQL 10.7 as the database backend for all applications.

\noindent{\bf Synthetic dataset.}
We use the tool described in \secref{sec:queryrewrites} to generate data 
for evaluation. 
We scale application data size to be 5--10GB (10K to 1M records per table), which is close to the size of data reported by application developers~\cite{dbsize1,dbsize2}.
\subsection{Constraint detection}
\seclabel{sec:eval-constraint-extract}
We report the total number of constraints detected by running \tool on each application in~\tabref{tab:constraint-count-source}.
\tool extracts an average of 289 constraints for each application.

\begin{table}[]
    \centering
    \caption{Number of different types of validation functions}
    \label{tab:val-stats}
    \resizebox{\linewidth}{!}{
        \begin{tabular}{c|c|c|c|c|c}
        \toprule
        Application & Builtin & Custom & Application & Built-in & Custom \\
        \midrule
        Redmine       & 277 (93\%)  &  22 (7\%)    & Dev.to      & 506 (91\%)    & 48 (9\%)\\
        Openproject   & 260 (87\%)  &  38 (13\%)   & Mastodon    & 256 (95\%)    & 14 (5\%)\\
        Openstreetmap & 274 (90\%)  &  4  (1\%)    & Spree       & 373 (94\%)    & 22 (6\%)\\
        \bottomrule
    \end{tabular}}
\end{table}

\noindent{\textbf{Missing constraints and Accuracy Analysis}
\tool extracts all constraints defined using built-in APIs and a subset of constraints defined in custom validators. As our approach is based on pattern matching, we are unable to enumerate every conceivable pattern defined in the user's custom validation code and thus can potentially miss constraints defined in custom validation functions. 
However, based on our observation of the 6 evaluated applications, as detailed in \cite{constropt-long}, over 85\% validation functions are defined using Rails' built-in validations and \tool already extracts a large portion of total constraints.
}

\begin{table}[t]
\centering
\caption{Number of model constraints extracted by analyzing application source code.}
\label{tab:constraint-count-source}
\vspace{-0.1in}
\scriptsize{
\resizebox{\columnwidth}{!}{
\begin{tabular}{m{1.7cm}|m{0.3cm}m{0.3cm}m{0.3cm}m{0.3cm}m{0.3cm}m{0.3cm}m{0.3cm}m{0.3cm}}
\toprule
Application                                      & Ds               & Dv               & Lm              & Lb              & Re               & Gi              & Op \\
\midrule
Data Validation                                  & 167              & 395              & 61              & 109             & {223}   & {853}  & {164} \\
\hline
Class Relations \newline and Field Definition    & {285}            & {111}   & {64}   & 48              & {89}    & {525}  & {103} \\
\hline
Total                                            & {452}            & {506}   & {125}  & {157}  & {312}   & {1378} & {267} \\
\toprule
\toprule
Application                                       & Sp               & Ro               & Da              & On              & Ma               & Ta              & Os \\
\midrule
Data Validation                                   & {129}   & {184}   & {64}   & {103}  & {108}   & 24              & {152} \\
\hline
Class Relations \newline and Field Definition     & {148}   & {103}   & {59}   & {65}   & {126}   & {21}   & {63} \\
\hline
Total                                             & {277}   & {287}   & {123}  & {168}  & {234}   & {45}   & {215} \\
\bottomrule
\end{tabular}
}
}
\end{table}

\begin{figure}
  \vspace{-0.2in}
  \centering
  \subfigure[Constraint defined in different applications]{
      \includegraphics[width=1.0\columnwidth]{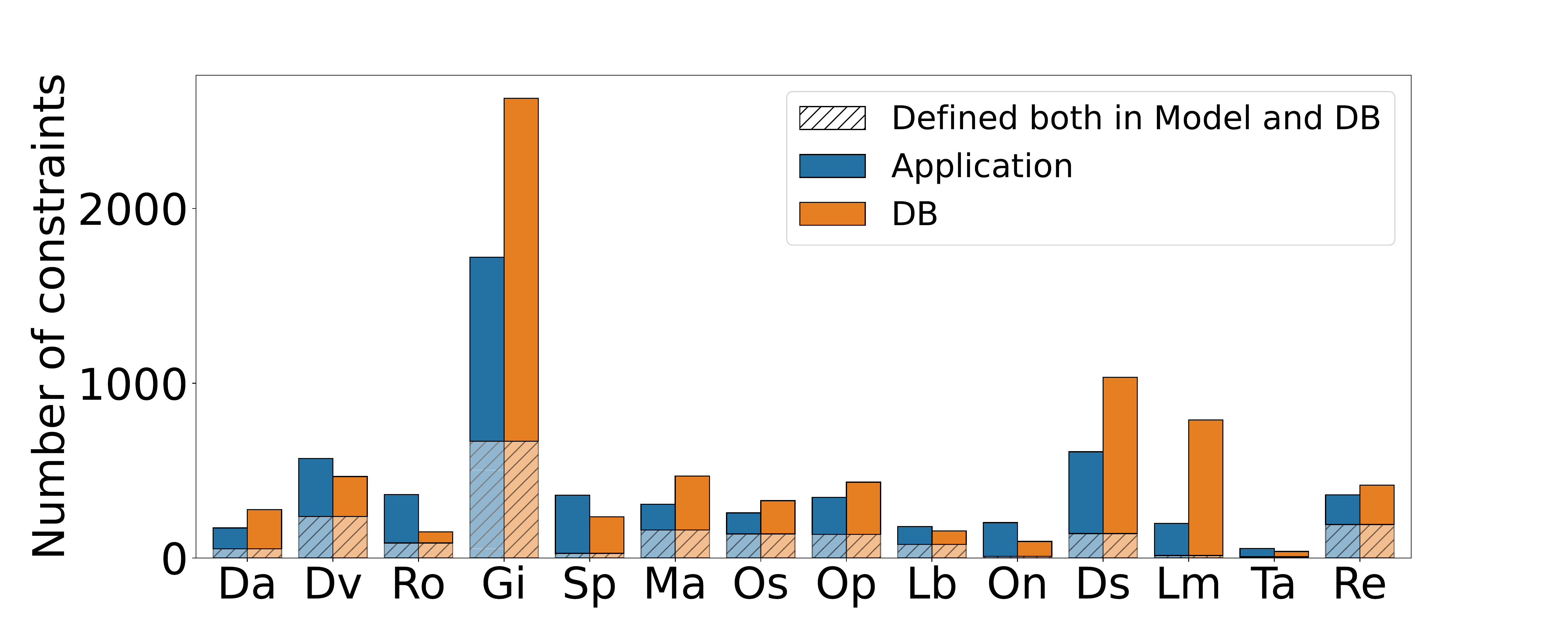}
      \label{fig:model-db-app}
  }
  \subfigure[Constraint type distribution. For both DB and application constraints, we sum the number across all 14 applications and group them by constraint type.]{
      \includegraphics[width=0.9\columnwidth]{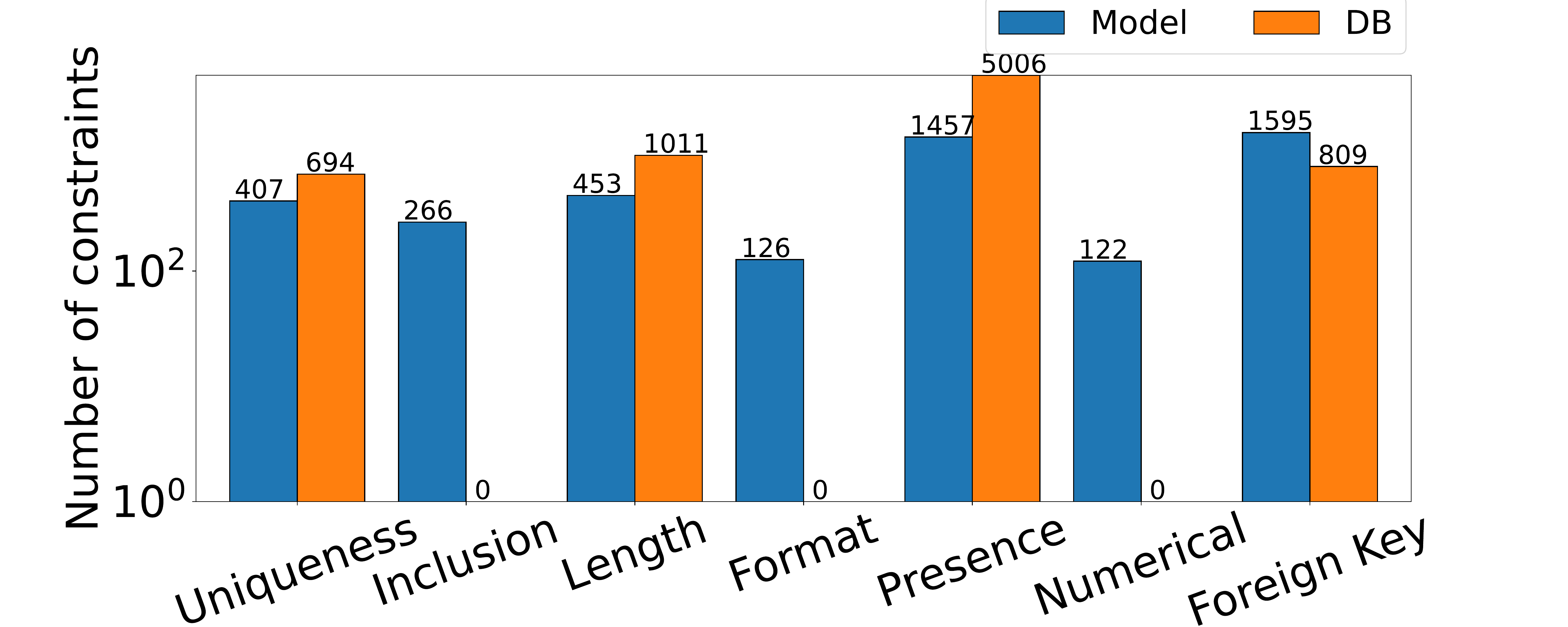}
      \label{fig:model-db-dis}
  }
\vspace{-0.2in}
\caption{Application and DB constraint comparison.}
\label{fig:fig:model-db-cmp}
\end{figure}

\noindent{\textbf{Comparison with database constraints.}} As described in \secref{sec:constraint-source}, developers can also explicitly declare database constraints in migration files. Constraints defined this way are installed as database constraints. Comparing the constraints extracted from model files and those installed in the database as shown in~\figref{fig:model-db-app}, 
there is a small percentage overlap which accounts for only 12\% of the total number of constraints. This demonstrates \tool's ability in discovering latent constraints defined only in the application code.

Moreover, as shown in~\figref{fig:model-db-dis}, the constraint type determines if it is defined in the application or the database.
For example, inclusion, format and numerical constraints are only defined in the application, as defining them in the database requires writing UDFs or as
{\code CHECK} constraints~\cite{constraint-psql}, which is tedious to implement.

\begin{table}
\centering
\caption{Compare with HyFD~\cite{papenbrock2016hybrid} and HyUCC~\cite{papenbrock2017hybrid}. ER = Error Rate. UCC = Unique Column Constraint. FD = Functional Dependency. Error rates for \tool are all 0\% because UCCs/FDs extracted by \tool are valid by construction. We show the total number of UCCs extracted by \tool and HyUCC, and the overlap defines the UCCs extracted by both methods. Similarly, we show the total number of FDs extracted by \tool and HyFD, as well as the overlap.}
\label{tab:cs-fd}
\resizebox{\linewidth}{!}{
    \begin{tabular}{c|c|c|c|c|}
    \toprule
    App 
    & \shortstack{Table name \\ (size in MB)}
    & \shortstack{\tool FD\\ /HyFD (ER)/overlap}
    & \shortstack{\tool UCC\\ /HyUCC (ER)/overlap}
    & \shortstack{HyFD/HyUCC/\\ \tool exec(s)} \\
    \midrule
    Dv   & notes                           (312)   & 7 / 29 (79\%) / 6       & 1 / 5 (80\%) / 1 & 3.06/2.94/0.06\\
    Re   & journal\textunderscore details  (703)   & 5 / 19 (74\%) / 5       & 1 / 4 (75\%) / 1 & 6.64/5.93/0.03\\
    Op   & auth\textunderscore sources     (316)   & 34 / 15 (60\%) / 10     & 2 / 5 (60\%) / 2 & 2.80/2.94/0.04\\
    Ma   & users                           (73)    & 70 / 74 (73\%) / 28     & 3 / 7 (57\%) / 4 & 3.21/2.65/0.08\\
    Os   & users                           (111)   & 584 / 1007 (44\%) / 551 & 4 / 34 (85\%)/ 4 & 3.80/3.31/0.07\\
    Sp   & spree\textunderscore assets     (2.9)   & 15  / 15 (60\%)         & 1 / 2 (50\%) / 1 & 0.44/0.45/0.09\\
    \bottomrule
    \end{tabular}
    }

\end{table}
\noindent{\textbf{Comparison with prior work.}}
We compare the correctness of extracted constraints and the execution time between \tool and two prior data-driven algorithms~\cite{metanome} on our synthetic data.
We run the HyUCC~\cite{papenbrock2017hybrid} discovery algorithm to detect unique columns and HyFD~\cite{papenbrock2016hybrid} to detect functional dependencies on the biggest synthetic table from each evaluated application.

As shown in ~\tabref{tab:cs-fd}, although HyUCC generates a superset of the UCCs that \tool generates, most constraints that are only detected by HyUCC are incorrect.
For instance, in Openstreetmap, HyUCC detects 34 unique columns, where only 5 of them are true keys when manually checking the application code. The other 29 detected columns appear to be unique looking at the data. For example, the {\code creation\_time} field of {\code User} is detected as a key as users are unlikely to be created at the same time. However, it is not a valid key as multiple users can theoretically be created concurrently. In contrast, \tool does not detect {\code creation\_time} as a key. It correctly identifies 5 true keys without false positives.
This illustrates a general issue with data-driven approaches to constraint discovery: while they can extract similar ones compared to \tool, the extracted constraints can be ephemeral due to the persistent data available at the time.

On extracting functional dependency constraints, We compare \tool with HyFD~\cite{papenbrock2016hybrid}. As shown in~\tabref{tab:cs-fd}, \tool and HyFD have some overlaps on detected functional dependencies (FDs), but \tool can detect FDs that HyFD missed. HyFD also reports FDs that are not detected by \tool, but upon manual inspection, the majority of such constraints are false positives. HyFD, like HyUCC, recognizes any column related to time information as a part of function dependencies. Also, since HyDB uses a fast approximation algorithm by only calculating from a subset of rows in the table, it falsely claims that many columns containing random values are unique (e.g., {\code encrypted\_password}).

Finally, \tool runs much faster than HyUCC and HyFD as it only scans the application source code once with complexity as described in ~\secref{sec:constraint-source} to extract unique columns and function dependencies concurrently. On the other hand, HyUCC/HyFD performs extraction by analyzing the data, which scales with the data size and is typically much bigger than the source code.

\subsection{Optimization opportunities}
\seclabel{sec:eval-opt-detect}
We next investigate \tool's ability to leverage constraints to optimize application performance on six applications, Redmine~\cite{redmine}, Dev.to~\cite{devto}, OpenProject~\cite{openproject}, Mastodon~\cite{mastodon}, OpenStreetmap~\cite{openstreetmap}, and Spree~\cite{spree}. We list the number of queries that use a constrained column in~\tabref{tab:cs-precheck-str2enum}. We first show the number of queries that can benefit from optimizing application code and physical design. We next measure the number of queries that can be rewritten using the \tool-detected constraints. 

\subsubsection{Precheck and optimizing physical design}
\begin{table}
\centering
\vspace{-0.1in}
\resizebox{\columnwidth}{!}{
    \begin{tabular}{c|c|c|c|c}
    \toprule
    Application 
    & \shortstack{Queries with\\constraints} 
    & \shortstack{Length\\Precheck} 
    & \shortstack{Format\\Precheck}   
    & \shortstack{Change Physical\\ Design (String to Enum)} \\
    \midrule
    Dev.to      & {4738}           & {676 (14.3\%)} & {456 (9.6\%)}  & {430 (9.1\%)}     \\
    Redmine     & {3511}           & {1270 (36.2\%)}& {302 (8.6\%)}  & {877 (25.0\%)}     \\
    OpenProject & {14845}          & {6329 (42.6\%)}& {502 (3.4\%)}  & {6373 (42.9\%)}    \\
    Mastodon    & {7059}           & {5469 (77.5\%)}& {5519 (78.2\%)}& {19 (0.3\%)}    \\
    Openstreetmap&{4889}           & {555 (11.4\%)} & 0 (0.0\%)               & {72 (1.5\%)}    \\
    Spree       & {4261}           & 15 {(0.4\%)}   & 0 (0.0\%)               & 185 {(4.3\%)}   \\
    \bottomrule
    \end{tabular}
}
\caption{Number of queries that can benefit from parameter precheck and changing data storage.}
\label{tab:cs-precheck-str2enum}
\end{table}

As shown in \tabref{tab:cs-precheck-str2enum},
the number of applicable optimizations in each category is determined by the number of constraints and queries that can benefit from the optimization. 
For example, despite having only 6 format constraints and 13 length constraints on user information columns (e.g., username, domain name), many queries in Mastodon are optimized, 
as 75\% of its queries search for username related records. 
On the other hand, as OpenStreetmap is a map application, its queries are almost exclusively about searching for location coordinates. 
However, there are only 3 format constraints and \revision{26} length constraints, and none of them are on coordinate related columns. 
Therefore, none of its queries benefit from the format precheck optimization.
Similarly, as Spree does not contain any format constraints, none of its queries benefit from the format precheck optimization either.

\subsubsection{Query Rewrite.}
\begin{figure}[t]
    \centering
    \includegraphics[width=0.7\columnwidth]{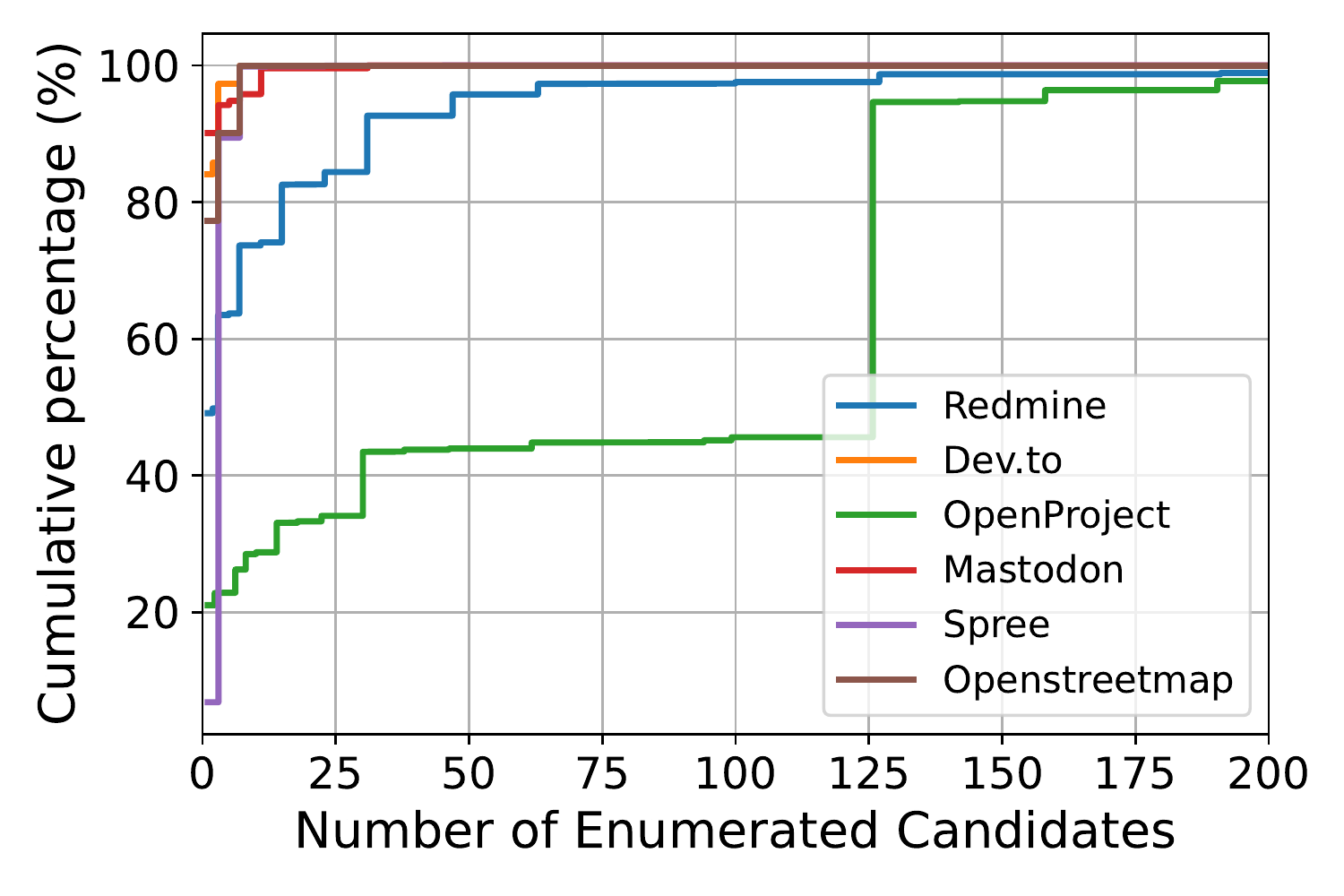}
    \vspace{-0.1in}
    \caption{{Enumeration Threshold Experiment}}
    \label{fig:rewrite-threshold}
    \vspace{-0.25in}
\end{figure}
\noindent {\textbf{Enumeration threshold.} 
Restricting the maximum number of rewrites enumerated can reduce enumeration time, at the expense of potentially not finding the optimal query rewrite. In \tool, we allow users to change the enumeration threshold based on their tolerance for enumeration time. To understand the impact of the enumeration threshold, we performed an empirical study on the evaluated apps by exhaustively enumerating every potential rewrite based on constraints and counting the number of candidate rewrites for each query template as shown in \figref{fig:rewrite-threshold}.}
{We observe that all queries from 4 of the 6 applications have fewer than 200 rewrite candidates, and only {22} out of {2032} Redmine queries and {239} of {14845} OpenProject queries have more than 200 candidate rewrites.
}
{The empirical study demonstrates that a small threshold can already discover the best rewrite for the majority of queries under most of the assessed apps, hence we recommend setting the enumeration threshold as a low value (e.g., 200).}

\begin{figure}[t]
    \centering
    \includegraphics[width=0.8\columnwidth]{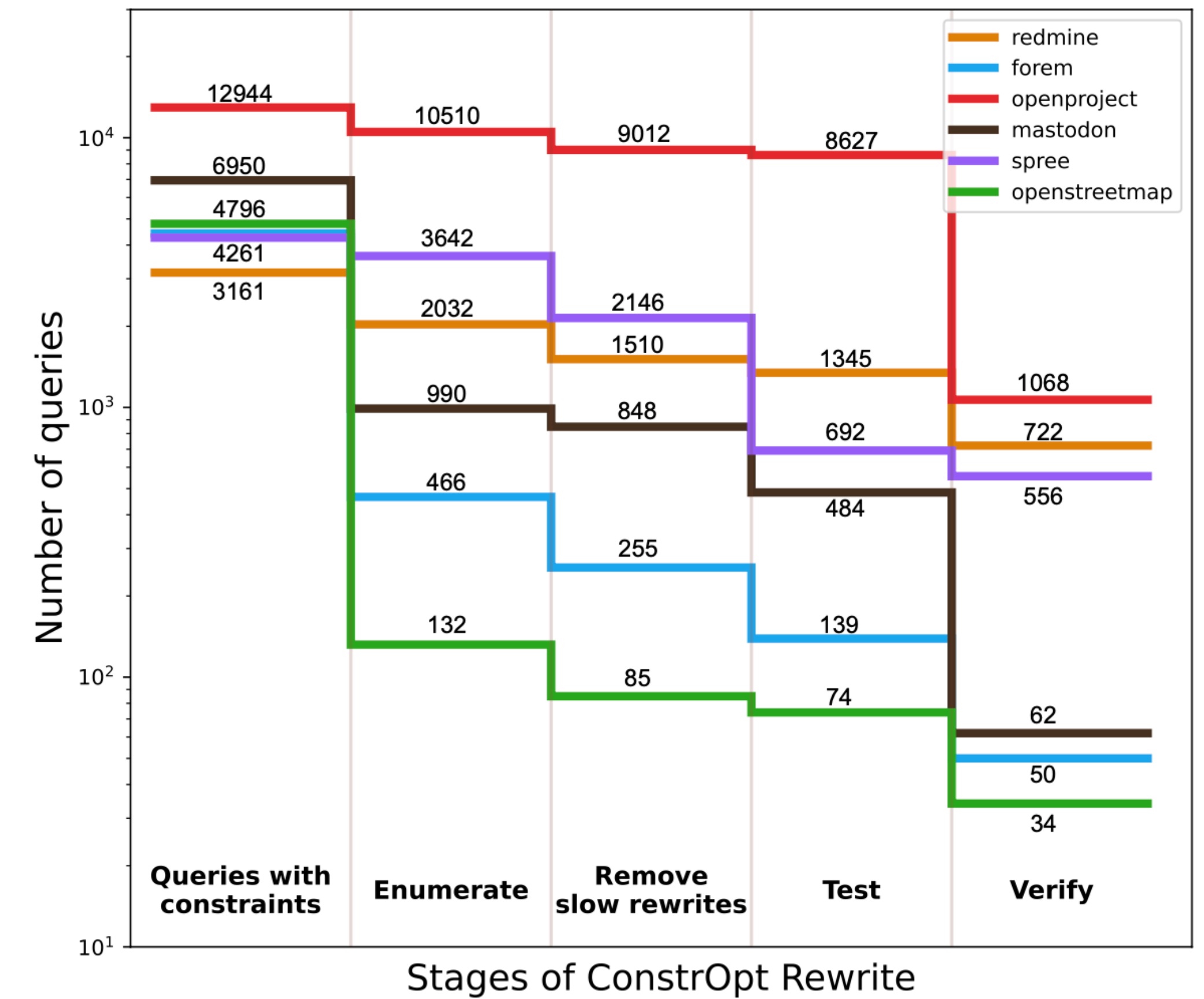}
    \caption{Number of queries after each rewrite step.}
    \label{fig:rewrite-stage-cnt}
\end{figure}

\noindent\textbf{Rewrite count.} In~\figref{fig:rewrite-stage-cnt}, we show the total number of queries rewritten by \tool, as well as the number of rewrites after each step. 
To count the number of queries with constraints, we extract all fields utilized by a query template and check if any of them contains a constraint.
By enumerating those queries, \tool generates candidate rewrites. 
The number of enumerated candidates varies across applications due to the varying styles of query templates used in each application.
For example, in Dev.to~\cite{devto}, we cannot apply the remove {\code DISTINCT} optimization since none of the query templates has the {\code DISTINCT} keyword. 
\tool then instantiates both the original query template and rewritten ones. \tool filters out the slow rewrites using the optimizer and retains those  with a lower cost. The last step before formal verification, test, continues to remove inequivalent queries by comparing query results on the sampled database, as explained in \secref{sec:queryrewrites}. Similarly, the estimating cost and testing steps are affected by query characteristics and vary by application. Since the first three steps filter away a considerable amount of the incorrect or slow rewrites, 1894 queries on average are finally sent to the verifier, which then proves that an average of over 415 queries can be optimized.

We examine the verified rewrites and find that some optimized queries  benefit from a combination of optimizations. For instance, \tool performs two optimizations on the query shown in~\lstref{eval:multi-opt} using two different constraints: the uniqueness leads to the removal of {\code DISTINCT}; and \tool changes the datatype of {\code users.type} from {\code varchar} to {\code enum} and performs selection on the transformed storage. Both optimizations lead to  2.4$\times$ speedup.


\begin{lstlisting} [language=SQL, caption={Two optimizations applied to a Redmine query.}, label={eval:multi-opt}, linewidth={9cm}]
--Constraints
--1.(members.user_id, members.project_id) pair is unique
--2.users.type can only take values from ['User', AnonymousUser']
--Query before
SELECT DISTINCT users.* FROM users INNER JOIN members ON members.user_id = users.id WHERE users.status = $1 AND (members.project_id = $2) AND users.type IN ($3)
--Query After (user.type has been changed to enumeration type)
SELECT users.* FROM users INNER JOIN members ON members.user_id = users.id WHERE users.status = $1 AND (members.project_id = $2) AND users.type IN ($3)
\end{lstlisting}

\begin{figure}
  \vspace{-0.1in}
  \centering
  \subfigure{
      \includegraphics[width=0.32\columnwidth]{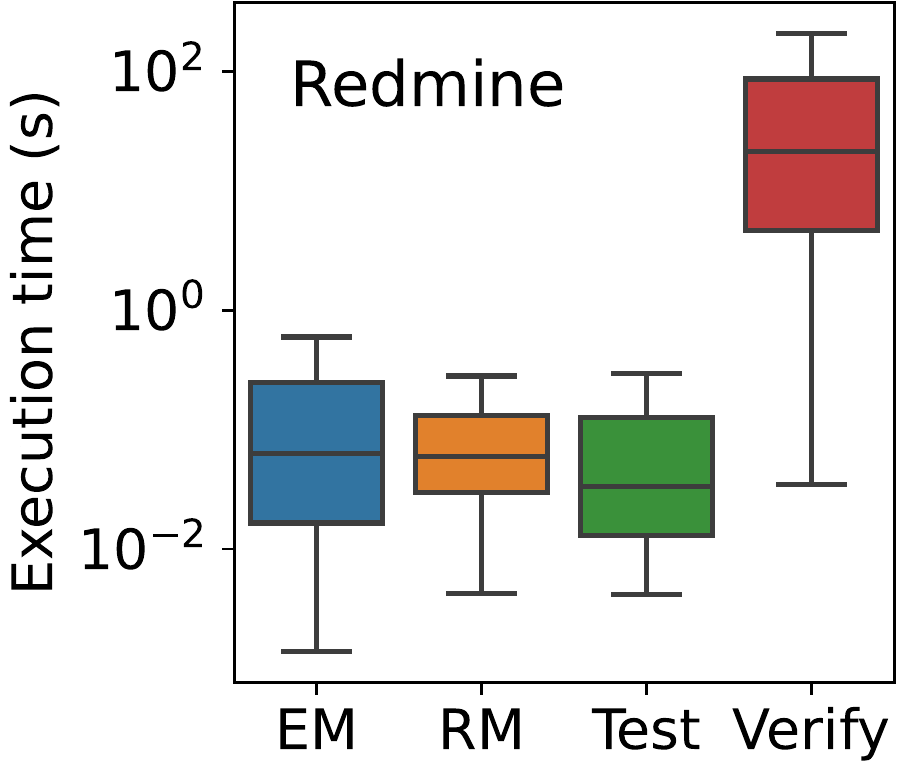}
      \label{fig:redmine-exec}
  }
  \subfigure{
      \includegraphics[width=0.3\columnwidth]{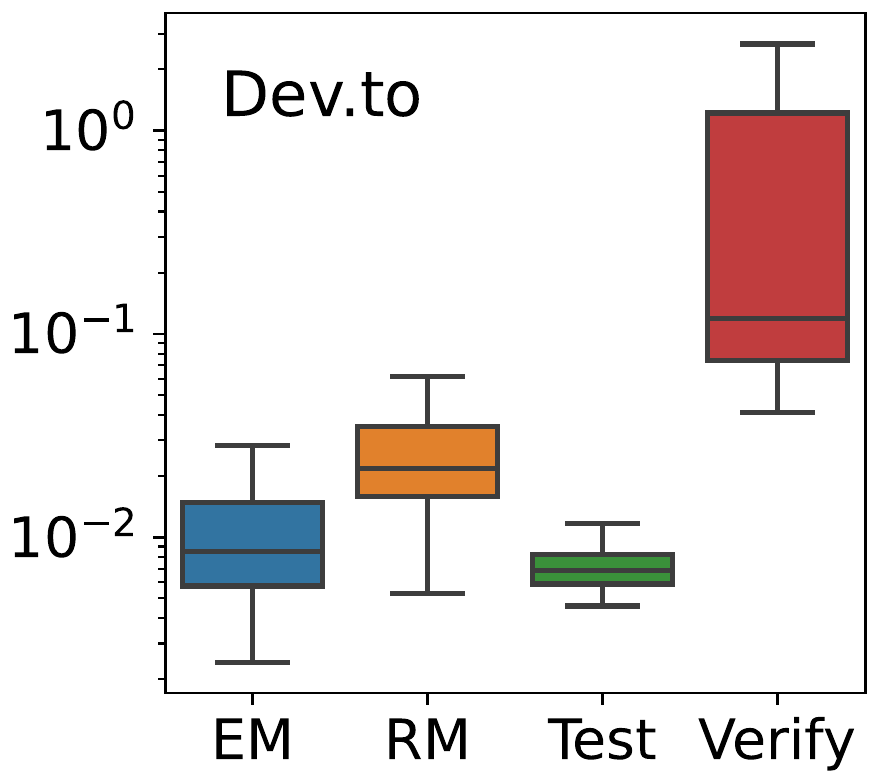}
      \label{fig:devto-exec}
  }
  \subfigure{
      \includegraphics[width=0.3\columnwidth]{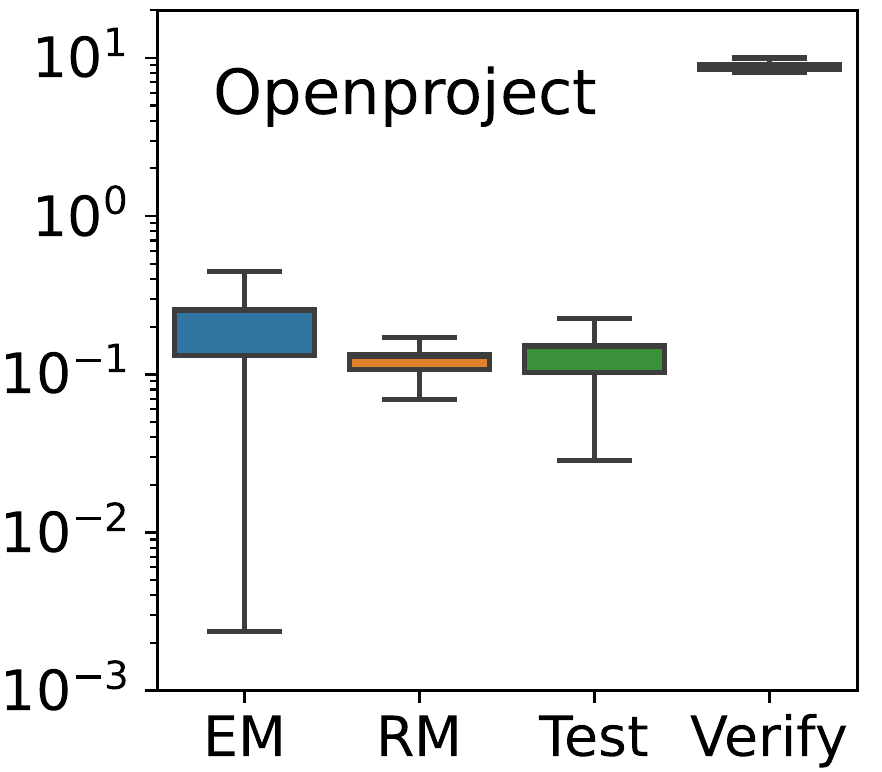}
      \label{fig:devto-exec}
  }
  \subfigure{
      \includegraphics[width=0.32\columnwidth]{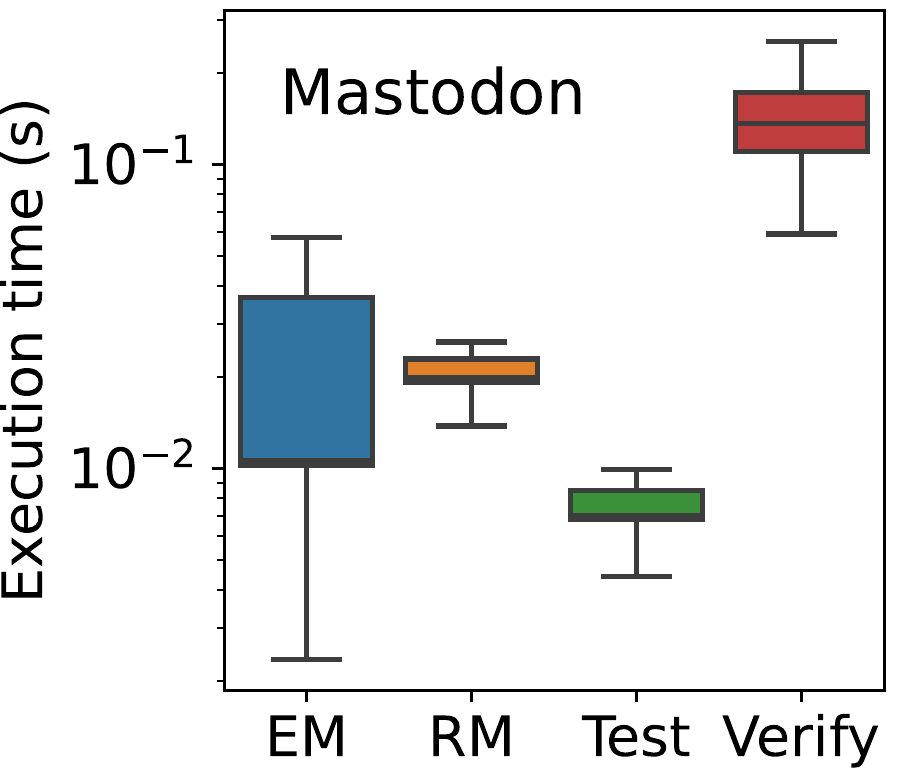}
      \label{fig:mastodon-exec}
  }
  \subfigure{
      \includegraphics[width=0.3\columnwidth]{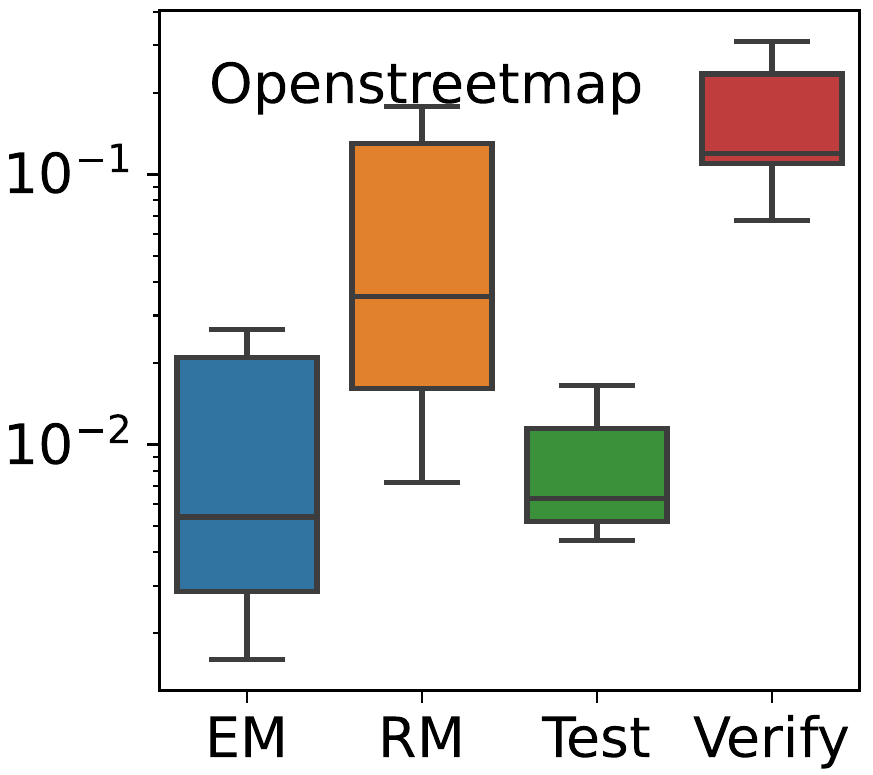}
      \label{fig:openstreetmap-exec}
  }
  \subfigure{
      \includegraphics[width=0.3\columnwidth]{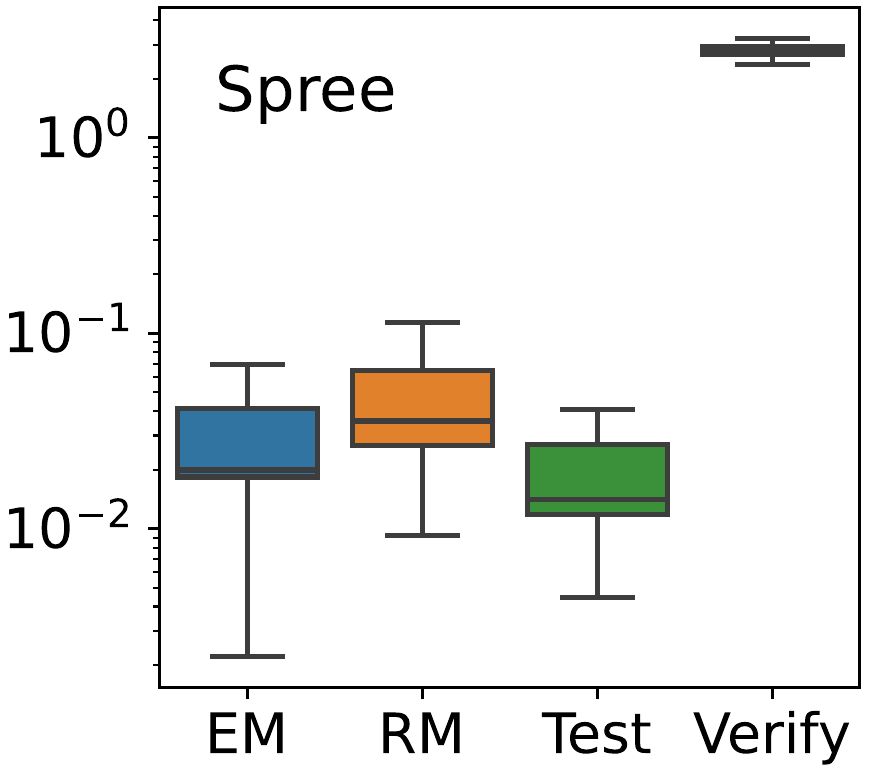}
      \label{fig:spree-exec}
  }
  \vspace{-0.3in}
  \caption{Average execution time for different rewrite steps of each query (EM: Enumerate; RM: Remove slow queries).}
  \label{fig:exec-time}
\end{figure}



\noindent\textbf{Execution time.}  The constraint extraction time for Redmine, OpenProject, Dev.to, Mastodon, Openstreetmap, and Spree are 1.30s, 1.41s, 0.55s, 0.49s, 0.62s, 0.27s, and 0.83s respectively.
The time breakdown of different steps is shown in~\figref{fig:exec-time}. For all apps, \tool completes the process in an acceptable amount of time. The average time to enumerate, remove slow rewrites, and run tests is less than 1s across all applications. Verification takes the longest time. However, even with Redmine and OpenProject, which have more complicated queries (with the longest query consisting of 1551 characters), the average verification time is still within 100s.
\subsection{Performance Evaluation}
\seclabel{sec:eval-perf}



\begin{figure}
  \centering
  \vspace{-0.15in}
  \subfigure{
      \includegraphics[width=0.31\columnwidth]{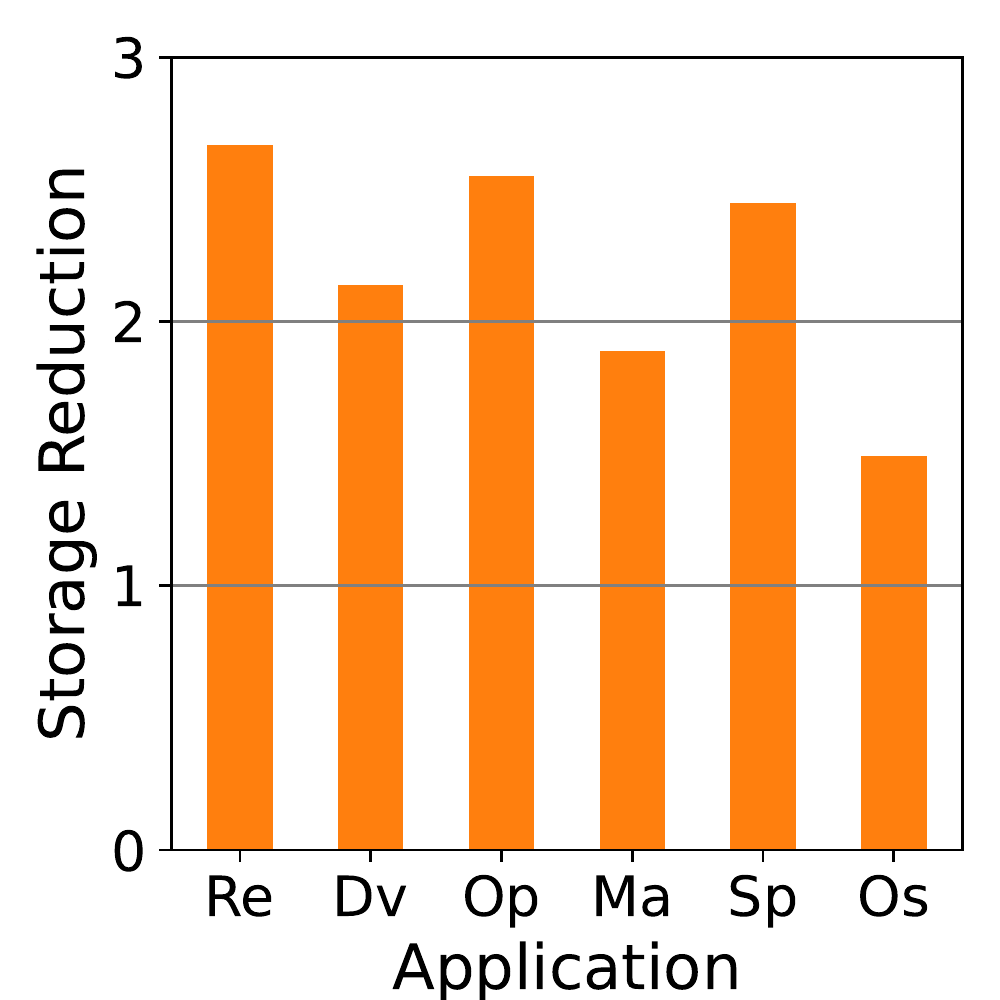}
      \label{fig:str2int_storage}
  }
  \subfigure{
      \includegraphics[width=0.31\columnwidth]{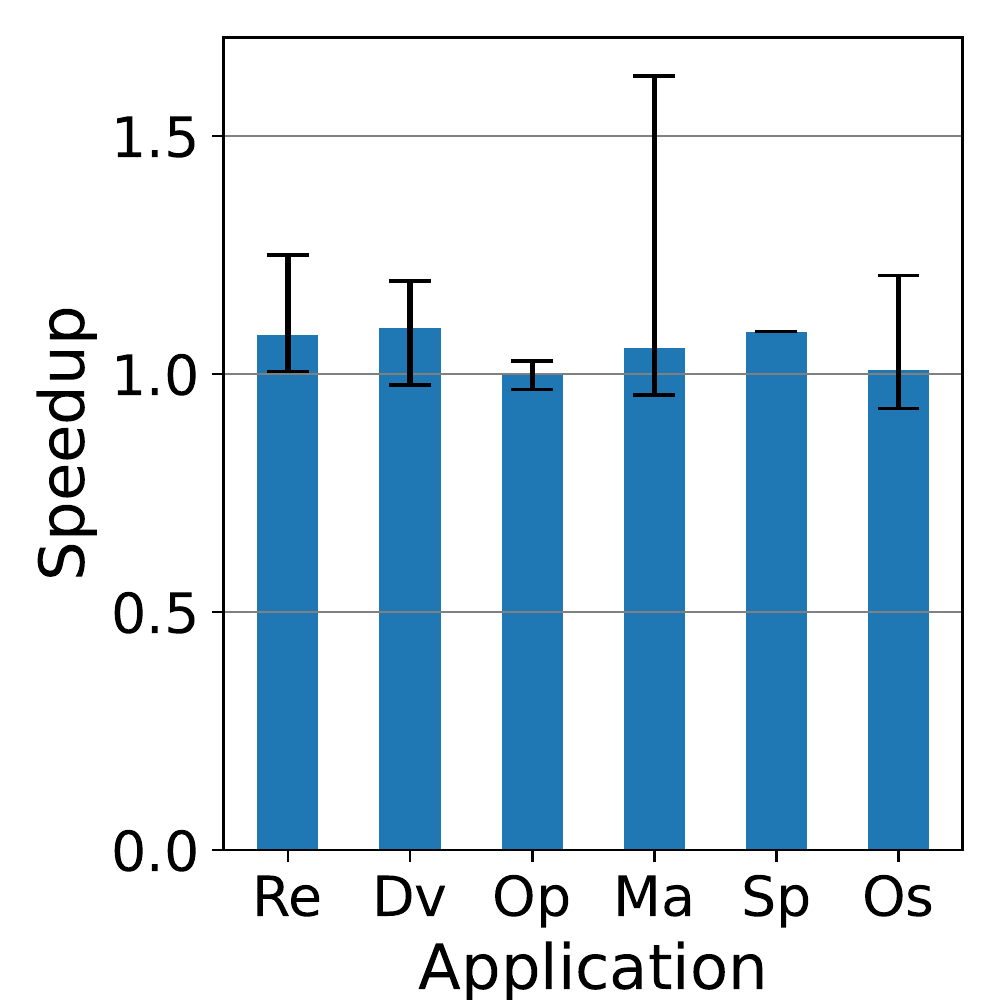}
      \label{fig:str2int_perf}
  }
  \subfigure{
      \includegraphics[width=0.3\columnwidth]{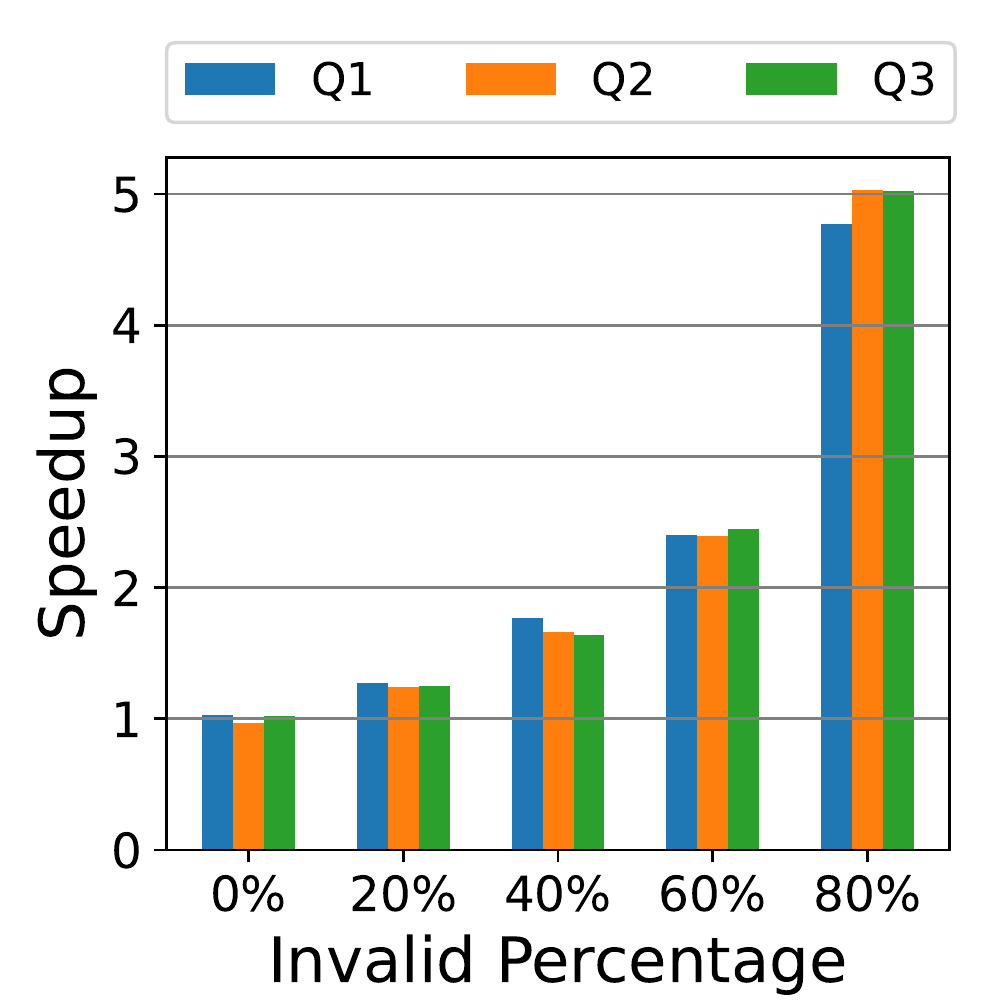}
      \label{fig:precheck-perf}
  }
  \vspace{-0.3in}
  \caption{Evaluation of optimizing code and data layout.}
  \label{fig:precheck-perf}
  \vspace{-0.2in}
\end{figure}

\begin{figure*}
  \centering
  \vspace{-0.15in}
  \subfigure{
      \includegraphics[width=0.64\columnwidth]{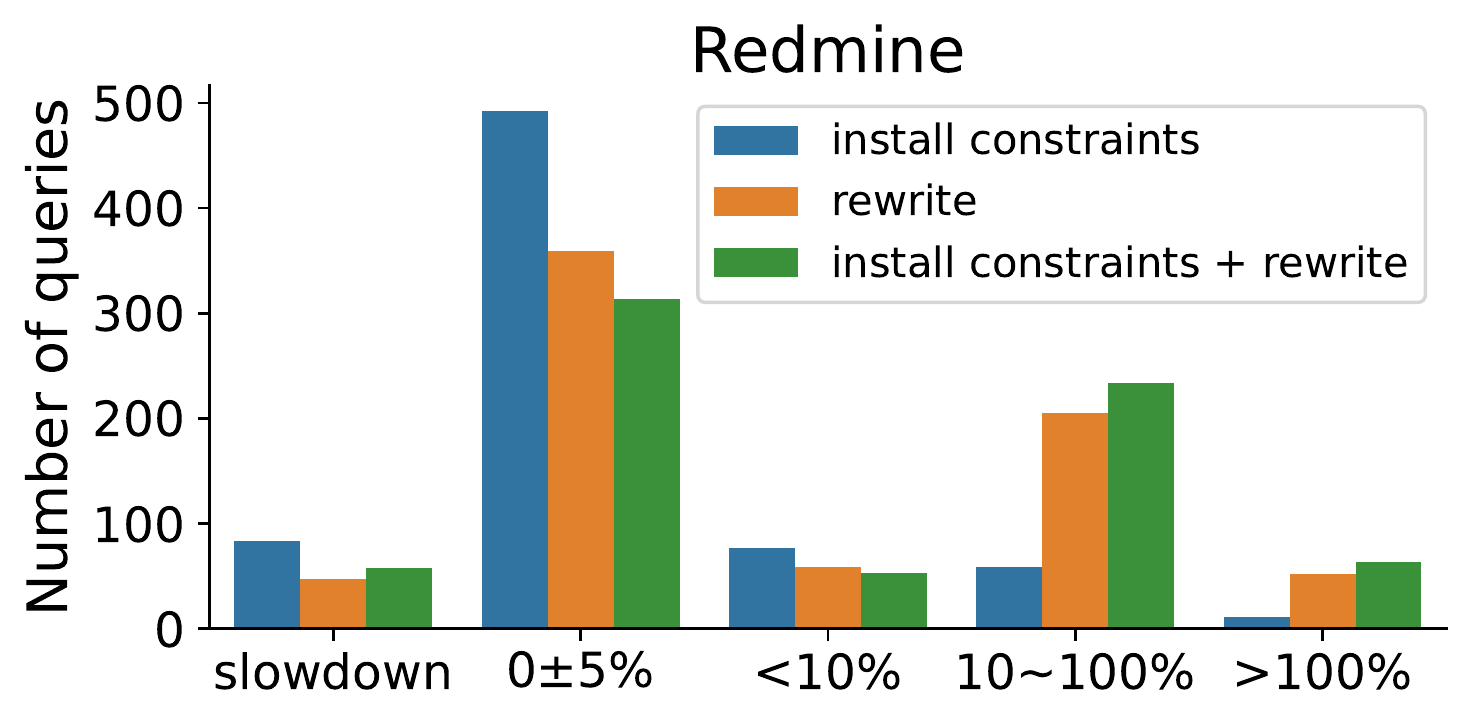}
      \label{fig:rewrite-redmine-perf}
  }
  \subfigure{
      \includegraphics[width=0.64\columnwidth]{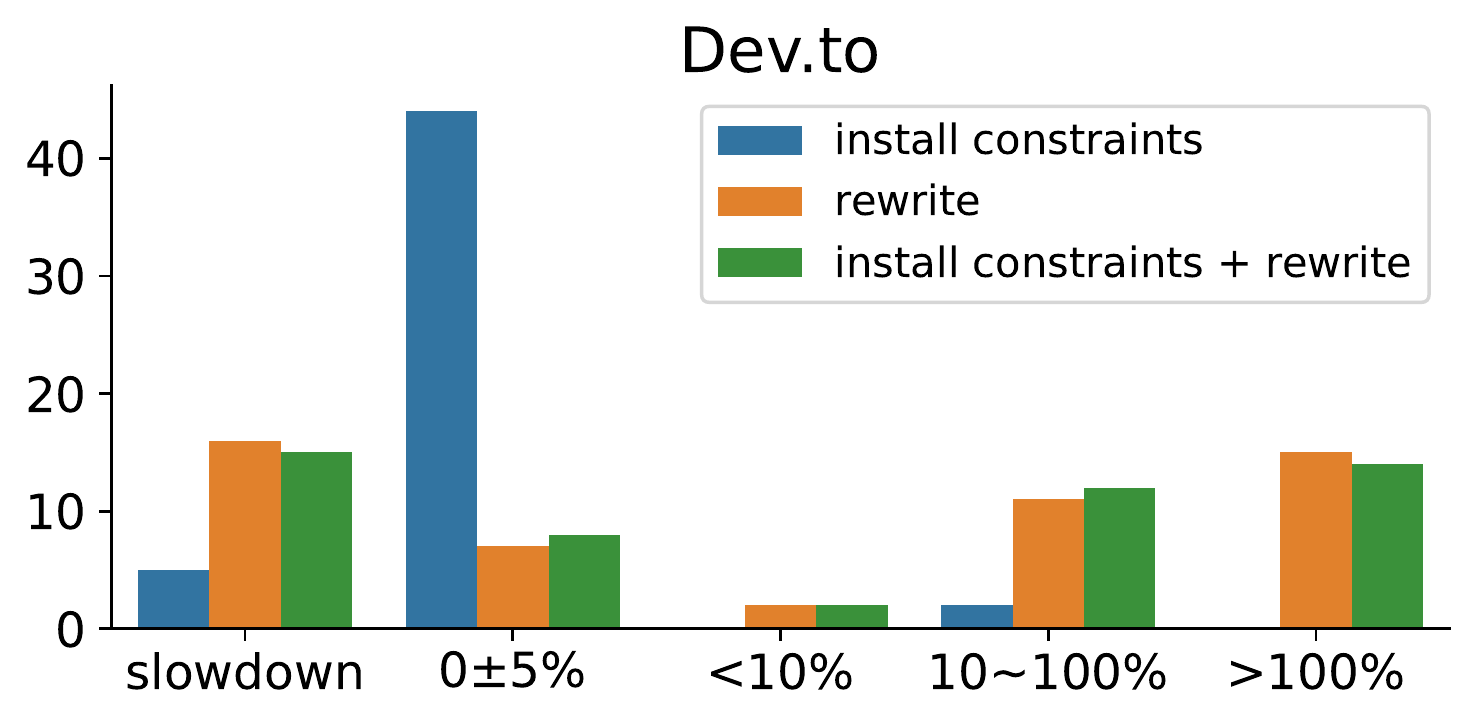}
      \label{fig:rewrite-devto-perf}
  }
  \subfigure{
      \includegraphics[width=0.64\columnwidth]{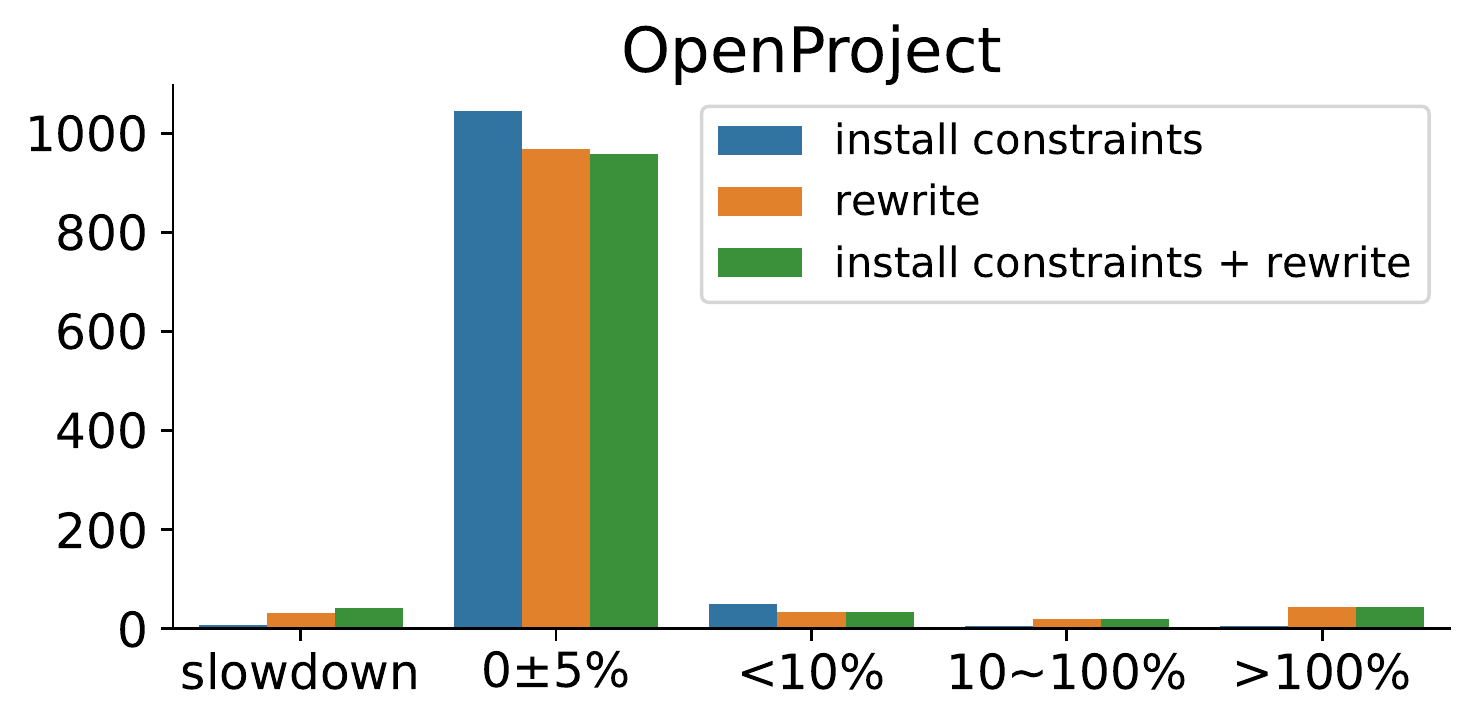}
      \label{fig:rewrite-openproject-perf}
  }
  \subfigure{
      \includegraphics[width=0.64\columnwidth]{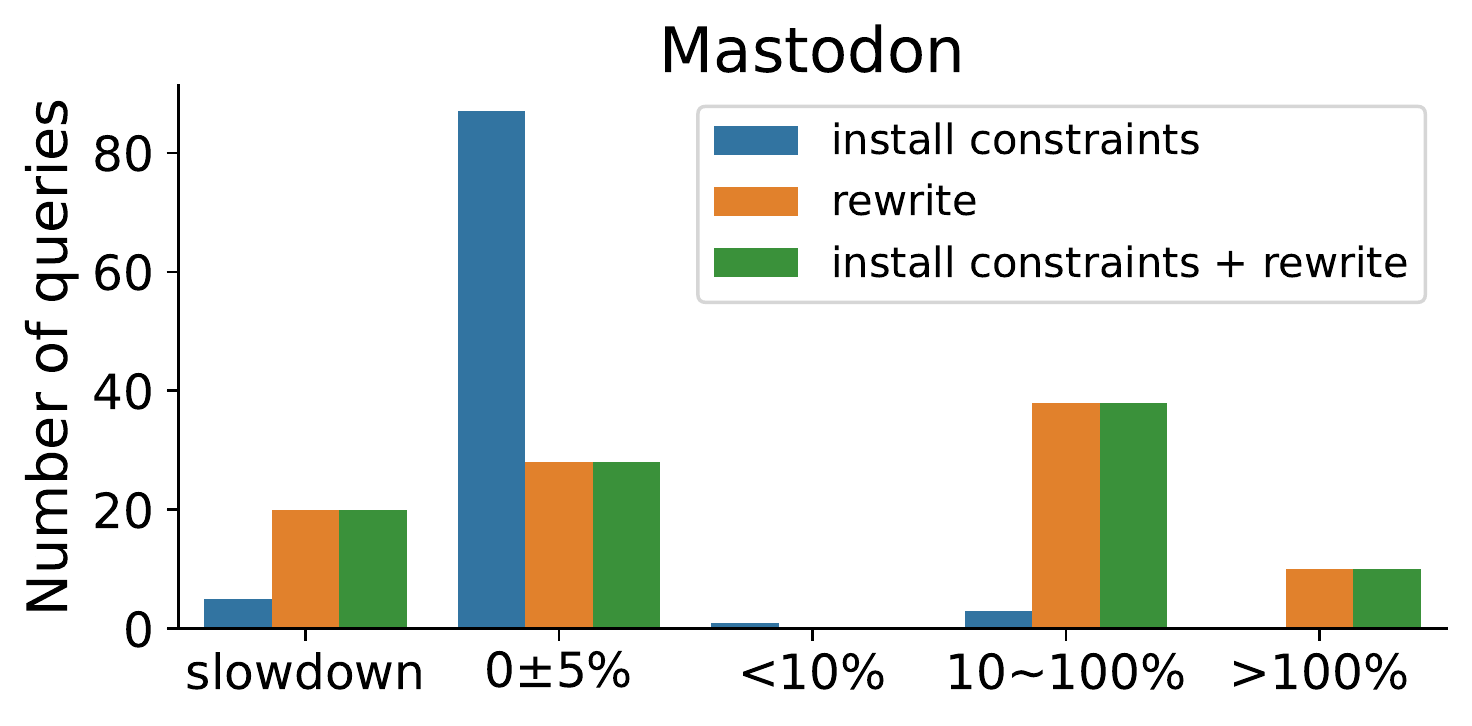}
      \label{fig:rewrite-mastodon-perf}
  }
  \subfigure{
      \includegraphics[width=0.64\columnwidth]{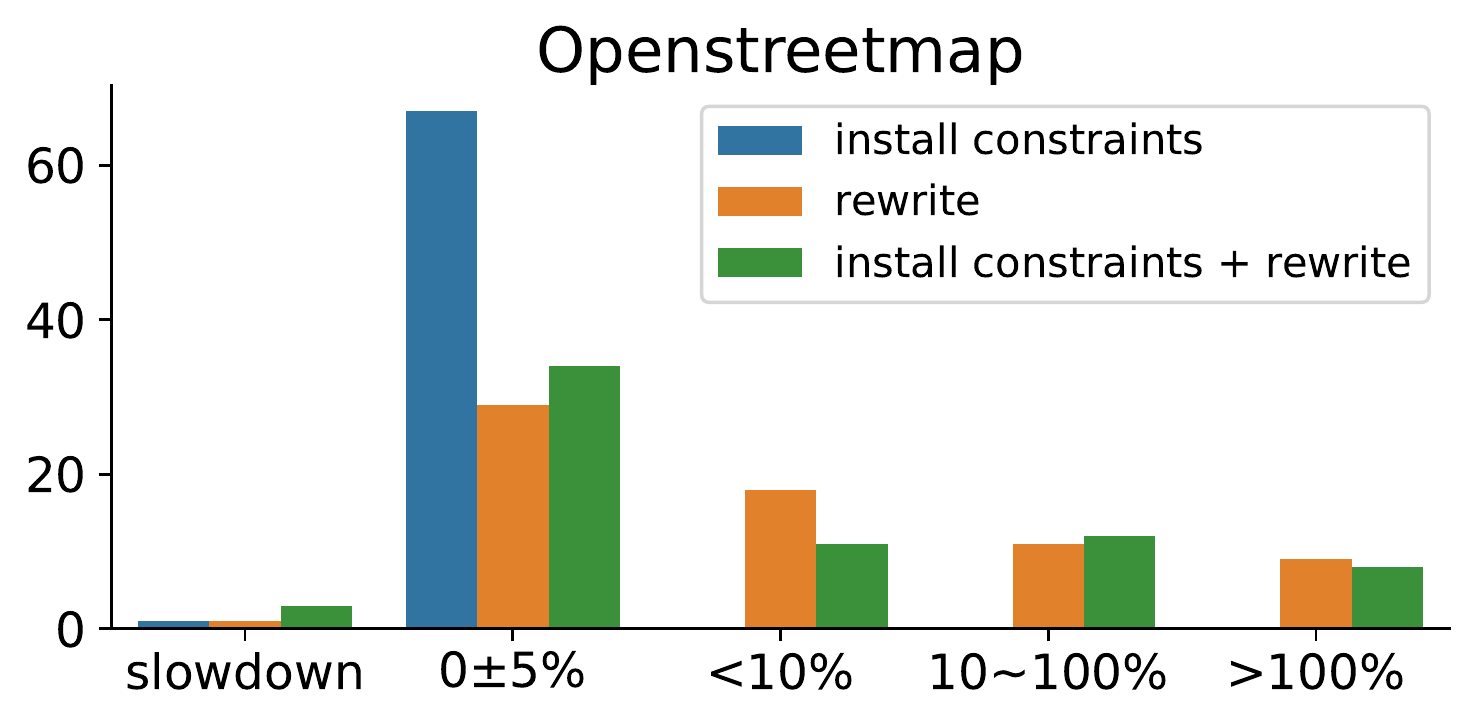}
      \label{fig:rewrite-openstreetmap-perf}
  }
  \subfigure{
      \includegraphics[width=0.64\columnwidth]{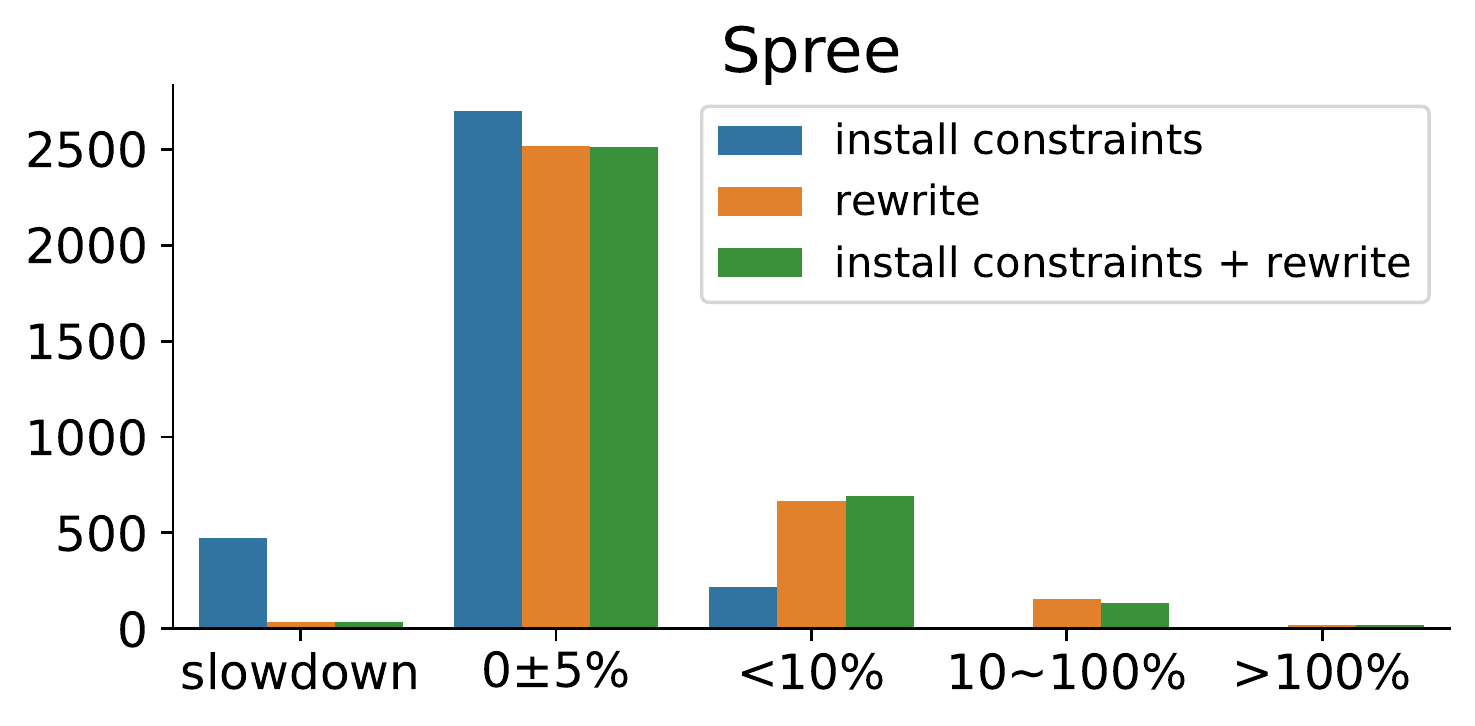}
      \label{fig:rewrite-spree-perf}
  }

\vspace{-0.2in}
\caption{Performance improvement after installing extracted constraints and performing rewrites. Improvement = ((execution time with only database constraints / execution time after optimization) - 1)$\times$ 100\%.}
\label{fig:rewrite-perf}
\vspace{-0.15in}
\end{figure*}

\begin{figure*}
  \centering
  \subfigure{
      \includegraphics[width=0.64\columnwidth]{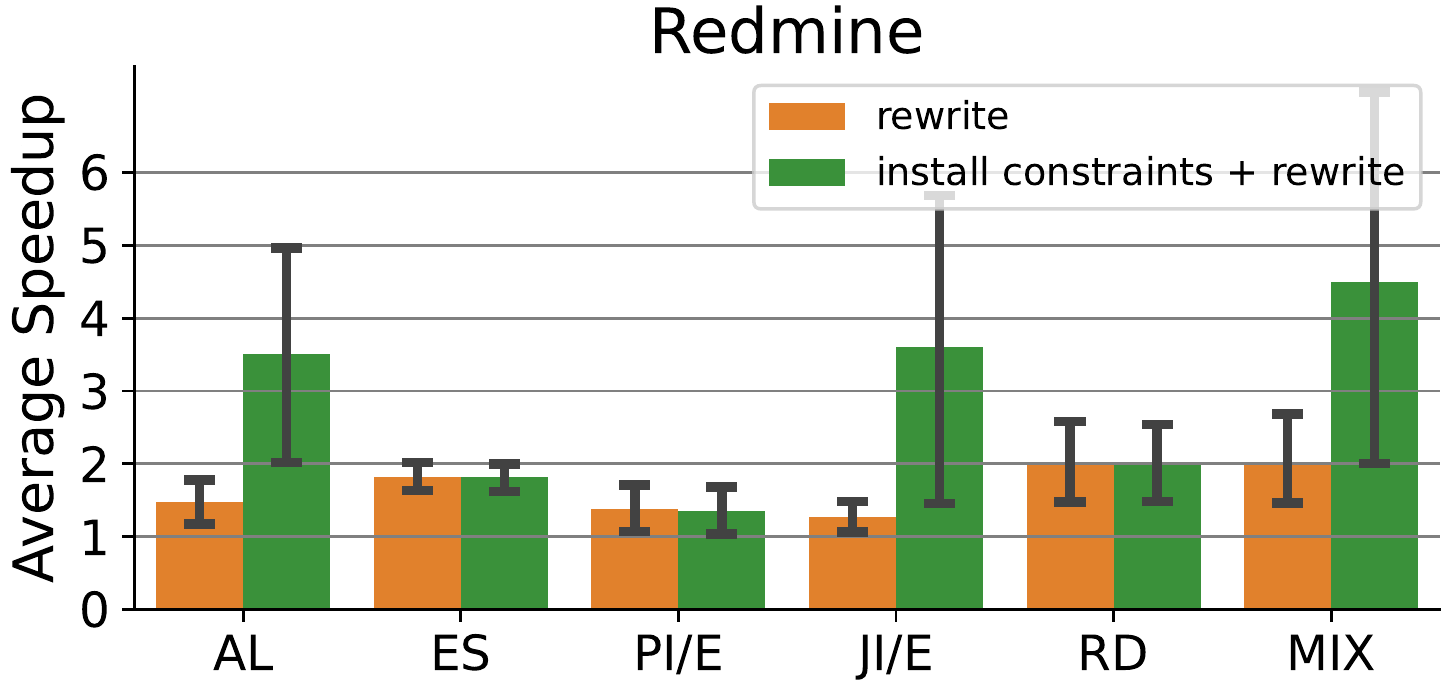}
      \label{fig:rewrite-type-redmine-perf}
  }
  \subfigure{
      \includegraphics[width=0.64\columnwidth]{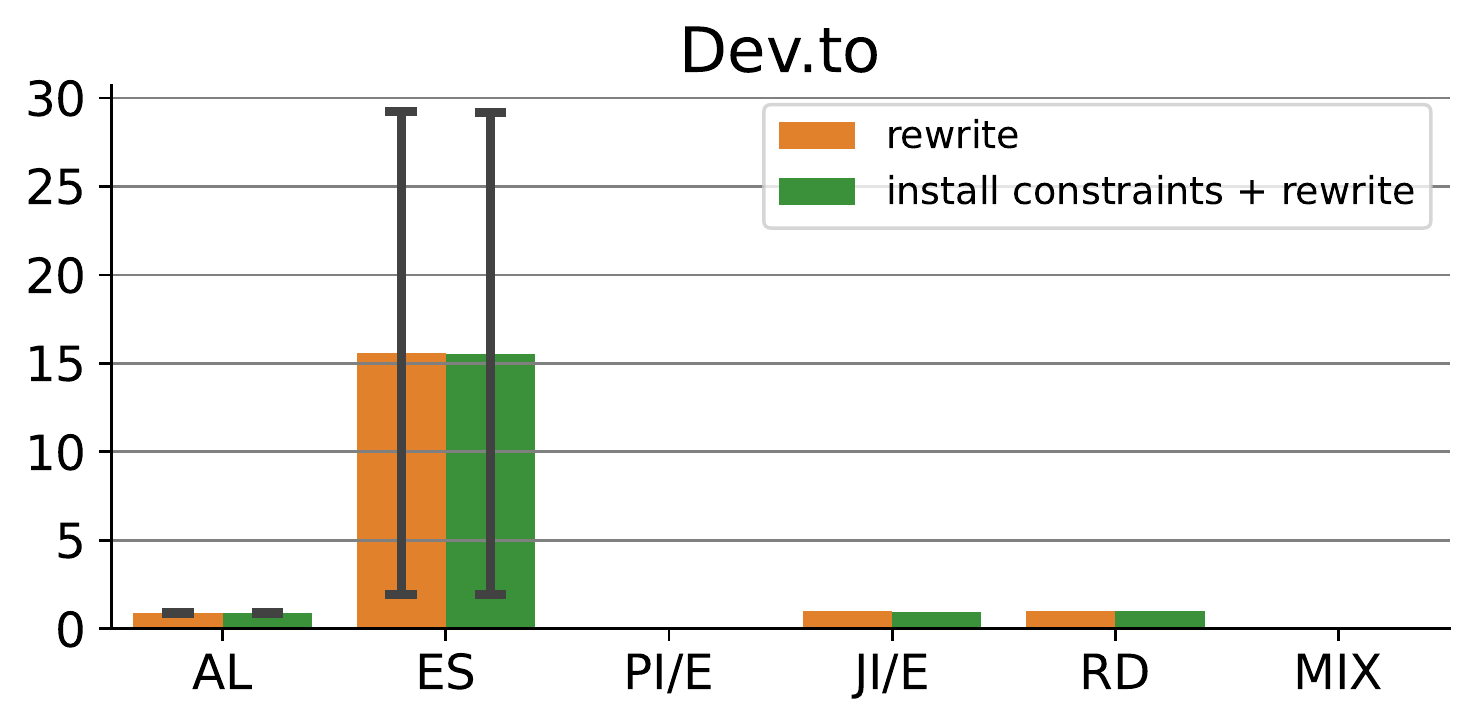}
      \label{fig:rewrite-type-devto-perf}
  }
  \subfigure{
      \includegraphics[width=0.64\columnwidth]{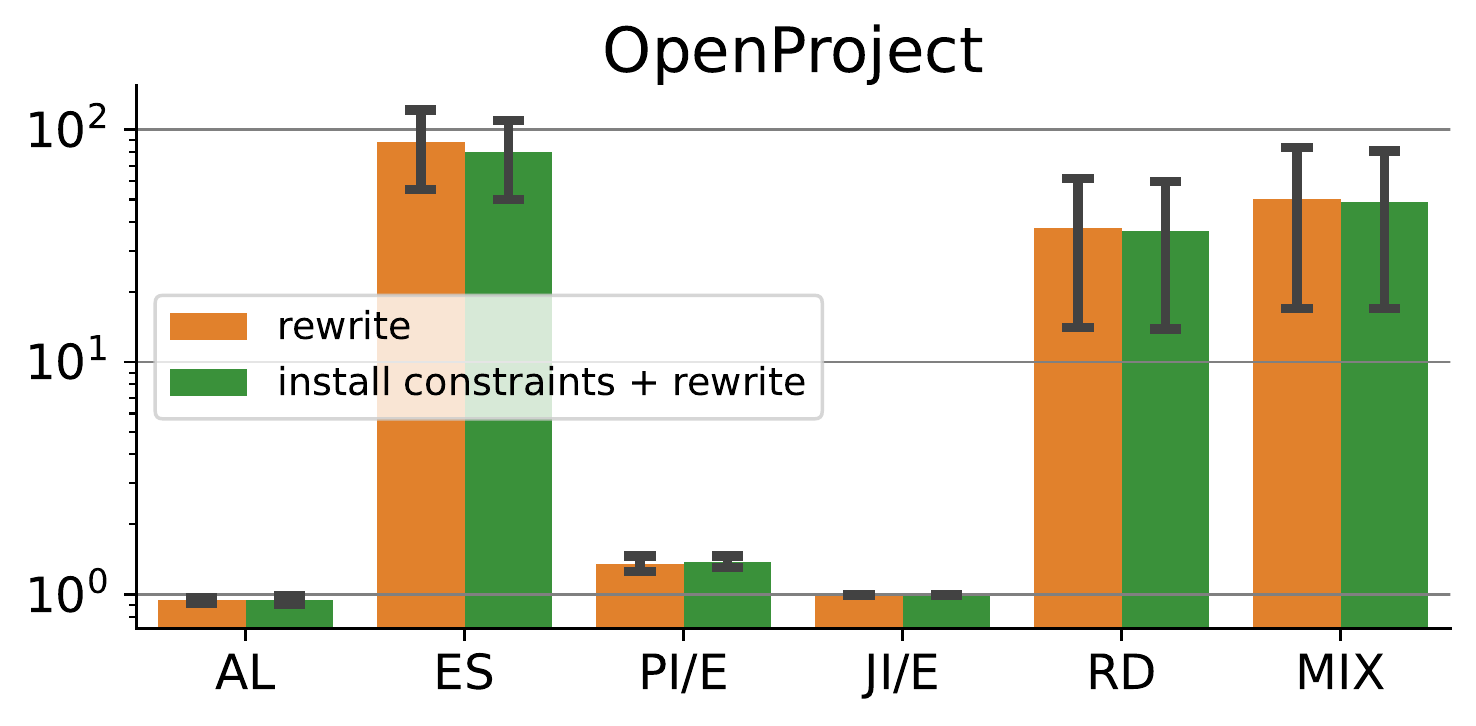}
      \label{fig:rewrite-type-openproject-perf}
  }
  \subfigure{
      \includegraphics[width=0.64\columnwidth]{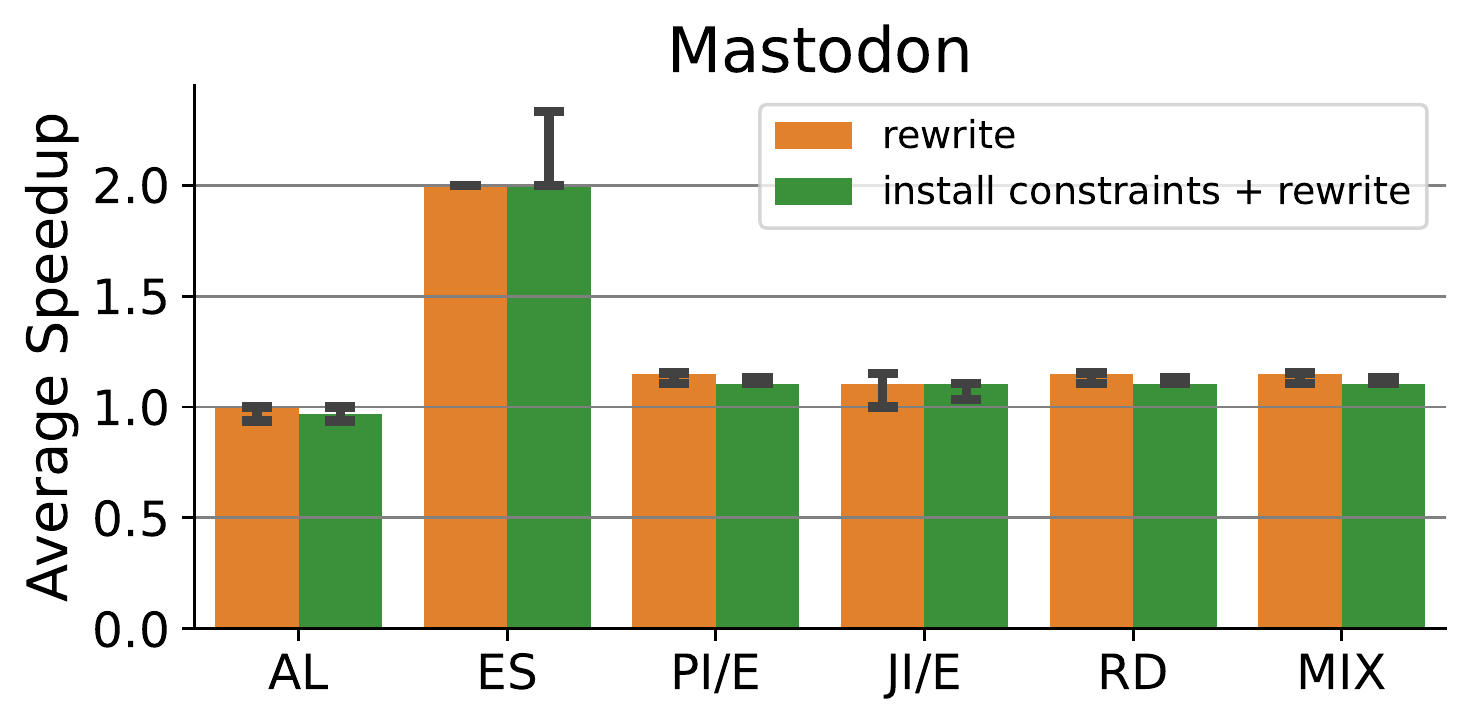}
      \label{fig:rewrite-type-redmine-perf}
  }
  \subfigure{
      \includegraphics[width=0.64\columnwidth]{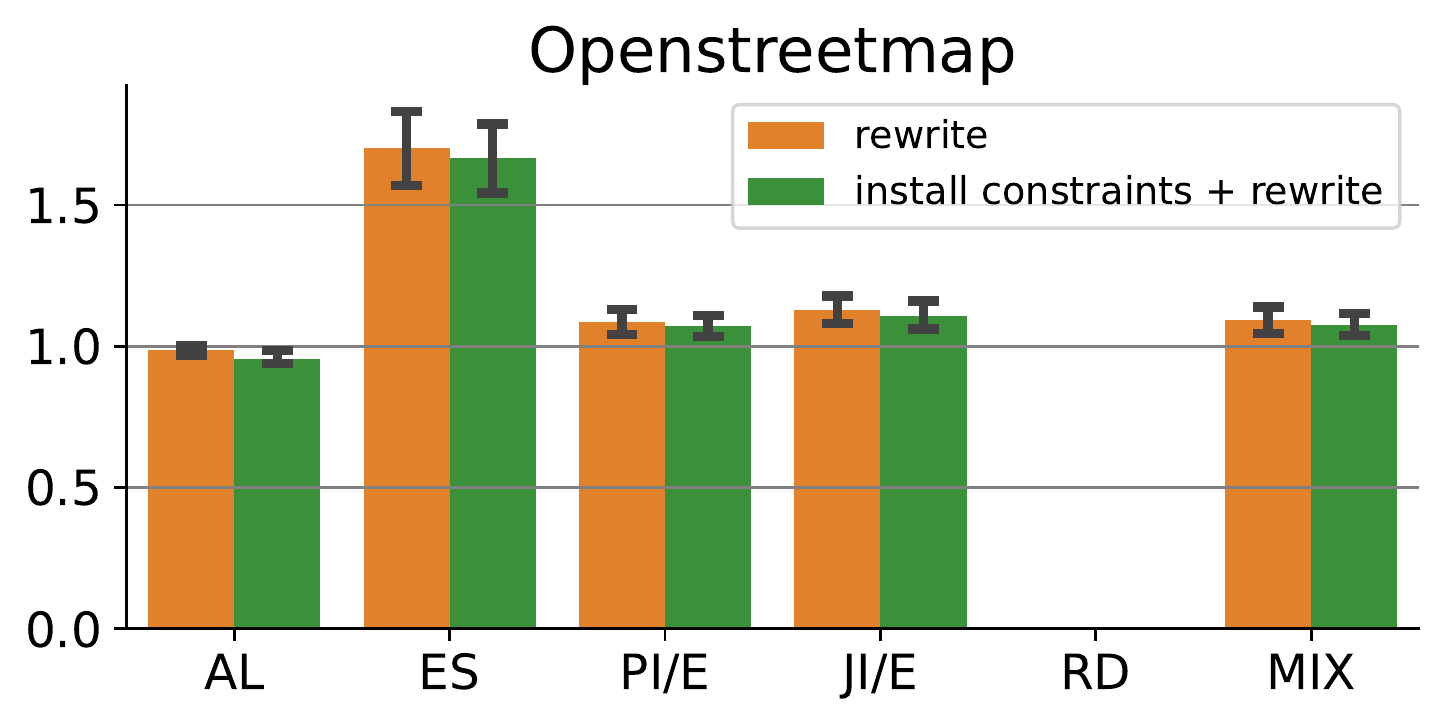}
      \label{fig:rewrite-type-devto-perf}
  }
  \subfigure{
      \includegraphics[width=0.64\columnwidth]{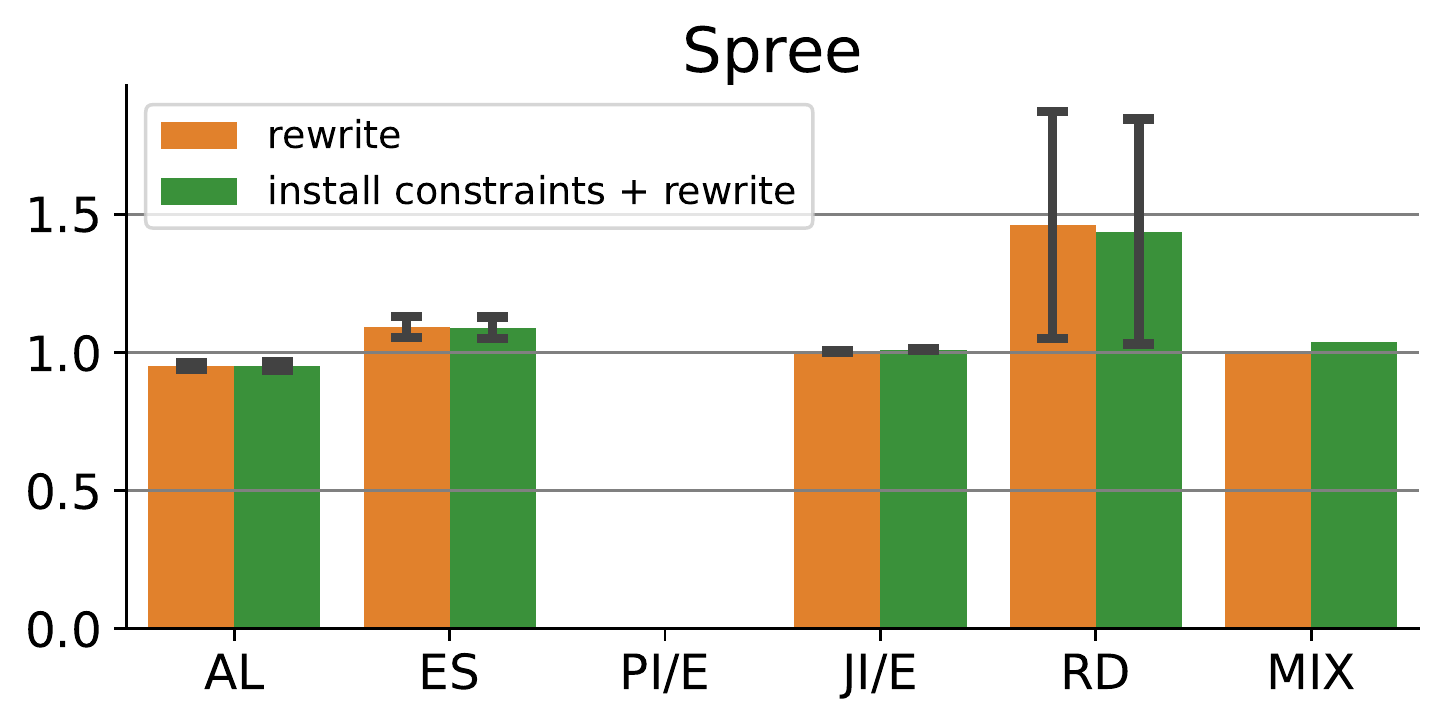}
      \label{fig:rewrite-type-openproject-perf}
  }
  
\vspace{-0.2in}
\caption{Speedup of various rewrite types. AL: Add Limit One, ES: Detecting Empty Set, PI/E: Introduce/Eliminate Predicate, JI/E: Introduce/Eliminate Join, RD: Remove Distinct, MIX: Mix of more than one rewrites. The error bar is the 25/75 percentile.}
\label{fig:rewrite-perf-type}
\vspace{-0.15in}
\end{figure*}

We first evaluate the storage reduction of changing data storage. For columns that can benefit from this optimization, assuming the candidate values are distributed evenly in our synthetic dataset, the average number of bytes for columns involved in the optimization are 10.70, 8.56, 10.21, 7.57, 5.97, 9.80 respectively. After changing to {\tt enum}, only 4 bytes are needed for each field, which reduced storage by
2.19$\times$ on average.
We also observe minor performance improvement by changing data storage. As shown in~\figref{fig:precheck-perf}, changing datatype from string to {\code enum} improves query performance by 1.05$\times$ across all applications.

The speedups of changing database schema are not significant as the majority of queries involve many columns, and columns with inclusion constraints only make up a small part of all those that are queried

To evaluate the benefits of adding prechecks on user inputs, we sample 3 queries and evaluate them under different ratios of invalid input. As shown in~\figref{fig:precheck-perf}, the speedup 
consistently increases as the percentage of invalid input increases. When 80\% of the input is invalid, we obtain the average speedup of 5$\times$. Meanwhile, 
adding precheck introduces negligible overhead.

We next evaluate the benefits of utilizing constraints to improve query performance. 
For each rewritten query, we first record the query execution time with only database constraints as the baseline. 
The time spent only doing the rewrite, only installing application constraints, 
and both installing application constraints and performing query rewriting are then recorded. 
We issue each query 30 times and average the results.
Note in \tool, the whole optimization process (constraint extraction, code optimization, data layout changes, and query rewrites) happens offline, as discussed in~\secref{sec:overview}.
Therefore, we only assess the query performance before and after rewrites to simulate the online setting.
\figref{fig:rewrite-perf} depicts the number of queries with various speed improvements.

\begin{lstlisting} [language=SQL, caption={Example of database optimizations with constraints from Redmine.}, label={eval:db-opt}, linewidth={9cm},escapeinside={}]
--Constraints, address is unique
--Original Query, creating a unique index on address speedups 12.3x
SELECT email_addresses.* FROM email_addresses WHERE email_addresses.address = 'bcfy@yaho.com' or user_id = 10;
\end{lstlisting}

As shown in~\figref{fig:rewrite-perf}, the database only optimizes a small fraction of queries. The largest speedup comes from creating an index on the unique column once the uniqueness constraints are installed. As demonstrated in~\lstref{eval:db-opt}, installing the uniqueness constraint on the {\code address} column creates an index, altering the query plan from a linear scan to an index lookup and getting 12.3$\times$ speedup.

We also examine the performance of just rewriting queries without installing the application constraints as shown in~\figref{fig:rewrite-perf}. The rewrite achieves comparable results as both installing and rewrite, which corresponds to our observation that the database underutilizes the installed constraints.
Lastly, we install the extracted constraints, utilize \tool to rewrite queries, and evaluate query performance. Among queries with constraints, over 7.2\% (52/722) for Redmine, 21.6\% (11/51) for Dev.to, 2.9\% (31/1086) for OpenProject, 12.9\% (8/62) for Mastodon, 17.6\% (6/34) for Openstreetmap, 1.8\% (10/556) for Spree have a speedup of more than 2$\times$. 
For some queries both installing constraints in the database and performing rewrites lead to more gains than only performing one of two. For example, in~\lstref{eval:db-opt}, installing the uniqueness constraint on the {\code address} column causes an index to be created. Subsequently, rewriting the query by adding {\code LIMIT 1} further speeds it up by 1.4$\times$ as it allows early return after finding the first matching record using the index.


We observe some slowdowns from the experiment and the causes are twofold. First, a few queries become slower after rewriting as the optimizer fails to predict the cost accurately, resulting in the rewritten query taking longer to execute. 
We believe the performance of those queries can still be improved with more accurate cost estimates. 
Second, many slowdowns are caused by our synthetically generated test dataset, which does not necessarily have the same data distribution as real-world data. This leads to some of the queries returning empty results (particularly for selections and joins), and 
any additional operation added as a result of \tool's rewrites causes a minor but obvious slowness. We believe our evaluation will be more accurate with real-world user data.

Lastly, we show the average speedup of different rewrites in ~\figref{fig:rewrite-perf-type}.
For all the applications, \tool increases the number of optimized queries across the different optimization categories. Detecting empty sets and deleting {\code DISTINCT} significantly reduce query latency.
The speedups mainly come from \tool’s ability to leverage constraints to simplify queries and save unnecessary computation to improve performance.
For example, when the selection result is known to be unique, \tool removes the DISTINCT operator from the query to avoid expensive sort or aggregate operations.
For Dev.to, Mastodon, and Openstreetmap, the majority of speedup comes from identifying empty sets, as queries in that application are short and straightforward (i.e., no joins and do not use {\code DISTINCT}) with little room for improvement. 
We also notice that the difference between only rewriting queries (orange bar) and both installing constraints and rewriting queries (green bar) is small. 
This is because the constraints are not fully utilized by the database, and the majority of the benefits can already be obtained by changing the SQLs. 
Redmine is an exception, where introducing constraints results in a higher speedup than merely changing the query. This is the same situation as the one previously discussed, where adding unique restrictions 
will result in an index being created on the unique column and speeding up data access.

\section{Related Work}



{\bf Constraint detection.} 
Techniques have been proposed to discover constraints from data.
\cite{huhtala1999tane,novelli2001fun,yao2008mining,abedjan2014dfd} extract dependencies by modeling the search space as a
power set lattice of attribute combinations and traverses it, while 
\cite{brown2003bhunt, ilyas2004cords, liu2016exploiting, kimura2009correlation} automatically discover soft and hard functional dependencies for big data query optimization. \tool instead detects constraints by code analysis. Compared to data-driven methods, \tool
scales well regardless of data size. It can also discover many other types of constraints in addition to functional and inclusion dependencies. 

\noindent{\bf Leveraging database constraints for query optimization.} 
Semantic query rewrite has been widely studied in the database literature. Some work provides theoretical results~\cite{FD-QO-theory,rewrite-theory0,rewrite-theory1,rewrite-theory2,rewrite-theory4}, while other work shows how this can be done in real systems~\cite{rewrite-system1, meier2013semantic,rewrite-system2}. All previous work uses heuristics~\cite{join-elimination,blog-disjunct} to leverage constraints, the contribution of \tool lies in combining different heuristics and automating the optimization process.

\section{Conclusion}
We presented \tool, a tool that extracts data constraints from applications to optimize queries. 
Our experiments show that \tool can discover many constraints from real-world applications, and can significantly improve application performance.
\clearpage
\bibliographystyle{ACM-Reference-Format}
\bibliography{ref}


\begin{thebibliography}{65}


\ifx \showCODEN    \undefined \def \showCODEN     #1{\unskip}     \fi
\ifx \showDOI      \undefined \def \showDOI       #1{#1}\fi
\ifx \showISBNx    \undefined \def \showISBNx     #1{\unskip}     \fi
\ifx \showISBNxiii \undefined \def \showISBNxiii  #1{\unskip}     \fi
\ifx \showISSN     \undefined \def \showISSN      #1{\unskip}     \fi
\ifx \showLCCN     \undefined \def \showLCCN      #1{\unskip}     \fi
\ifx \shownote     \undefined \def \shownote      #1{#1}          \fi
\ifx \showarticletitle \undefined \def \showarticletitle #1{#1}   \fi
\ifx \showURL      \undefined \def \showURL       {\relax}        \fi
\providecommand\bibfield[2]{#2}
\providecommand\bibinfo[2]{#2}
\providecommand\natexlab[1]{#1}
\providecommand\showeprint[2][]{arXiv:#2}

\bibitem[\protect\citeauthoryear{??}{dja}{[n.d.]a}]%
        {django-inheritance}
 \bibinfo{year}{[n.d.]}\natexlab{a}.
\newblock \bibinfo{title}{Class inheritance in Django}.
\newblock
  \bibinfo{howpublished}{\url{https://docs.djangoproject.com/en/3.1/topics/db/models/\#model-inheritance}}.
\newblock


\bibitem[\protect\citeauthoryear{??}{con}{[n.d.]}]%
        {constraint-psql}
 \bibinfo{year}{[n.d.]}\natexlab{}.
\newblock \bibinfo{title}{Constraint type in PostgreSQL}.
\newblock
  \bibinfo{howpublished}{\url{https://www.postgresql.org/docs/13/ddl-constraints.html}}.
\newblock


\bibitem[\protect\citeauthoryear{??}{dev}{[n.d.]}]%
        {devto}
 \bibinfo{year}{[n.d.]}\natexlab{}.
\newblock \bibinfo{title}{Dev.to}.
\newblock
\newblock
\newblock
\shownote{An open source software for building communities.
  \url{https://github.com/forem/forem}.}


\bibitem[\protect\citeauthoryear{??}{dia}{[n.d.]}]%
        {diaspora}
 \bibinfo{year}{[n.d.]}\natexlab{}.
\newblock \bibinfo{title}{Diaspora}.
\newblock
\newblock
\newblock
\shownote{\url{https://github.com/diaspora/diaspora}.}


\bibitem[\protect\citeauthoryear{??}{dis}{[n.d.]}]%
        {discourse}
 \bibinfo{year}{[n.d.]}\natexlab{}.
\newblock \bibinfo{title}{Discourse}.
\newblock \bibinfo{howpublished}{\url{https://github.com/discourse/discourse}}.
\newblock


\bibitem[\protect\citeauthoryear{??}{dbs}{[n.d.]a}]%
        {dbsize1}
 \bibinfo{year}{[n.d.]}\natexlab{a}.
\newblock \bibinfo{title}{Discourse Database size}.
\newblock \bibinfo{howpublished}{\url{https://meta.discourse.org/t/88941/}}.
\newblock


\bibitem[\protect\citeauthoryear{??}{dbs}{[n.d.]b}]%
        {dbsize2}
 \bibinfo{year}{[n.d.]}\natexlab{b}.
\newblock \bibinfo{title}{Discourse Database size}.
\newblock \bibinfo{howpublished}{\url{https://meta.discourse.org/t/57598/}}.
\newblock


\bibitem[\protect\citeauthoryear{??}{blo}{[n.d.]}]%
        {blog-disjunct}
 \bibinfo{year}{[n.d.]}\natexlab{}.
\newblock \bibinfo{title}{Disjunctive subquery optimization}.
\newblock
  \bibinfo{howpublished}{\url{https://nenadnoveljic.com/blog/disjunctive-subquery-optimization/}}.
\newblock


\bibitem[\protect\citeauthoryear{??}{dja}{[n.d.]b}]%
        {django-validation}
 \bibinfo{year}{[n.d.]}\natexlab{b}.
\newblock \bibinfo{title}{Django validation API}.
\newblock
  \bibinfo{howpublished}{\url{https://docs.djangoproject.com/en/3.0/ref/models/fields/validators}}.
\newblock


\bibitem[\protect\citeauthoryear{??}{enu}{[n.d.]a}]%
        {enum-mysql}
 \bibinfo{year}{[n.d.]}\natexlab{a}.
\newblock \bibinfo{title}{Enumeration type in MySQL}.
\newblock
  \bibinfo{howpublished}{\url{https://dev.mysql.com/doc/refman/8.0/en/enum.html}}.
\newblock


\bibitem[\protect\citeauthoryear{??}{enu}{[n.d.]b}]%
        {enum-psql}
 \bibinfo{year}{[n.d.]}\natexlab{b}.
\newblock \bibinfo{title}{Enumeration type in PostgreSQL}.
\newblock
  \bibinfo{howpublished}{\url{https://www.postgresql.org/docs/9.1/datatype-enum.html}}.
\newblock


\bibitem[\protect\citeauthoryear{??}{git}{[n.d.]}]%
        {gitlab}
 \bibinfo{year}{[n.d.]}\natexlab{}.
\newblock \bibinfo{title}{Gitlab}.
\newblock
\newblock
\newblock
\shownote{\url{https://github.com/gitlabhq/gitlabhq}.}


\bibitem[\protect\citeauthoryear{??}{joi}{[n.d.]}]%
        {join-elimination}
 \bibinfo{year}{[n.d.]}\natexlab{}.
\newblock \bibinfo{title}{Join elimination}.
\newblock
  \bibinfo{howpublished}{\url{https://blog.jooq.org/2017/09/01/join-elimination-an-essential-optimiser-feature-for-advanced-sql-usage/}}.
\newblock


\bibitem[\protect\citeauthoryear{??}{lob}{[n.d.]}]%
        {lobsters}
 \bibinfo{year}{[n.d.]}\natexlab{}.
\newblock \bibinfo{title}{Lobsters}.
\newblock
\newblock
\newblock
\shownote{\url{https://github.com/lobsters/lobsters}.}


\bibitem[\protect\citeauthoryear{??}{loo}{[n.d.]}]%
        {loomio}
 \bibinfo{year}{[n.d.]}\natexlab{}.
\newblock \bibinfo{title}{Loomio}.
\newblock
\newblock
\newblock
\shownote{\url{https://github.com/loomio/loomio}.}


\bibitem[\protect\citeauthoryear{??}{mas}{[n.d.]}]%
        {mastodon}
 \bibinfo{year}{[n.d.]}\natexlab{}.
\newblock \bibinfo{title}{Mastodon}.
\newblock
\newblock
\newblock
\shownote{\url{https://github.com/mastodon/mastodon}.}


\bibitem[\protect\citeauthoryear{??}{met}{[n.d.]}]%
        {metanome}
 \bibinfo{year}{[n.d.]}\natexlab{}.
\newblock \bibinfo{title}{Metanome}.
\newblock \bibinfo{howpublished}{\url{https://bit.ly/metanome}}.
\newblock


\bibitem[\protect\citeauthoryear{??}{one}{[n.d.]}]%
        {onebody}
 \bibinfo{year}{[n.d.]}\natexlab{}.
\newblock \bibinfo{title}{Onebody}.
\newblock
\newblock
\newblock
\shownote{A social networking website to help management for churches.
  \url{https://github.com/seven1m/onebody}.}


\bibitem[\protect\citeauthoryear{??}{ope}{[n.d.]a}]%
        {openproject}
 \bibinfo{year}{[n.d.]}\natexlab{a}.
\newblock \bibinfo{title}{OpenProject, A web-based project management
  software.}
\newblock \bibinfo{howpublished}{\url{https://github.com/opf/openproject}}.
\newblock


\bibitem[\protect\citeauthoryear{??}{ope}{[n.d.]b}]%
        {openstreetmap}
 \bibinfo{year}{[n.d.]}\natexlab{b}.
\newblock \bibinfo{title}{OpenStreetMap}.
\newblock
\newblock
\newblock
\shownote{\url{https://github.com/openstreetmap/openstreetmap-website}.}


\bibitem[\protect\citeauthoryear{??}{mig}{[n.d.]}]%
        {migration}
 \bibinfo{year}{[n.d.]}\natexlab{}.
\newblock \bibinfo{title}{Rails migration files}.
\newblock
  \bibinfo{howpublished}{\url{https://guides.rubyonrails.org/active_record_migrations.html}}.
\newblock


\bibitem[\protect\citeauthoryear{??}{rai}{[n.d.]a}]%
        {rails-inheritance}
 \bibinfo{year}{[n.d.]}\natexlab{a}.
\newblock \bibinfo{title}{Rails single table inheritance}.
\newblock
  \bibinfo{howpublished}{\url{https://api.rubyonrails.org/classes/ActiveRecord/Inheritance.html}}.
\newblock


\bibitem[\protect\citeauthoryear{??}{sta}{[n.d.]a}]%
        {state-machine1}
 \bibinfo{year}{[n.d.]}\natexlab{a}.
\newblock \bibinfo{title}{Rails state machine library}.
\newblock \bibinfo{howpublished}{\url{https://github.com/aasm/aasm}}.
\newblock


\bibitem[\protect\citeauthoryear{??}{rai}{[n.d.]b}]%
        {rails-validation}
 \bibinfo{year}{[n.d.]}\natexlab{b}.
\newblock \bibinfo{title}{Rails validation API}.
\newblock
  \bibinfo{howpublished}{\url{https://guides.rubyonrails.org/active_record_validations.html}}.
\newblock


\bibitem[\protect\citeauthoryear{??}{red}{[n.d.]}]%
        {redmine}
 \bibinfo{year}{[n.d.]}\natexlab{}.
\newblock \bibinfo{title}{Redmine}.
\newblock
\newblock
\newblock
\shownote{A project management application.
  \url{https://github.com/redmine/redmine}.}


\bibitem[\protect\citeauthoryear{??}{pg-}{[n.d.]}]%
        {pg-remove-distinct}
 \bibinfo{year}{[n.d.]}\natexlab{}.
\newblock \bibinfo{title}{Remove Distinct in PostgreSQL}.
\newblock
  \bibinfo{howpublished}{\url{https://commitfest.postgresql.org/35/2433/}}.
\newblock


\bibitem[\protect\citeauthoryear{??}{ror}{[n.d.]}]%
        {ror-ecommerce}
 \bibinfo{year}{[n.d.]}\natexlab{}.
\newblock \bibinfo{title}{Ror ecommerce}.
\newblock
\newblock
\newblock
\shownote{\url{https://github.com/drhenner/ror_ecommerce}.}


\bibitem[\protect\citeauthoryear{??}{rai}{[n.d.]c}]%
        {rails}
 \bibinfo{year}{[n.d.]}\natexlab{c}.
\newblock \bibinfo{title}{Ruby on Rails, a ruby web application framework}.
\newblock \bibinfo{howpublished}{\url{https://rubyonrails.org/}}.
\newblock


\bibitem[\protect\citeauthoryear{??}{z3}{[n.d.]}]%
        {z3}
 \bibinfo{year}{[n.d.]}\natexlab{}.
\newblock \bibinfo{title}{Solver z3}.
\newblock \bibinfo{howpublished}{\url{https://pypi.org/project/z3-solver/}}.
\newblock


\bibitem[\protect\citeauthoryear{??}{spr}{[n.d.]}]%
        {spree}
 \bibinfo{year}{[n.d.]}\natexlab{}.
\newblock \bibinfo{title}{Spree}.
\newblock
\newblock
\newblock
\shownote{An ecommerce application. \url{https://github.com/spree/spree/}.}


\bibitem[\protect\citeauthoryear{??}{sta}{[n.d.]b}]%
        {statemachinedoc}
 \bibinfo{year}{[n.d.]}\natexlab{b}.
\newblock \bibinfo{title}{State Machine API Document}.
\newblock
  \bibinfo{howpublished}{\url{https://www.rubydoc.info/github/pluginaweek/state_machine/StateMachine/MacroMethods}}.
\newblock


\bibitem[\protect\citeauthoryear{??}{tra}{[n.d.]}]%
        {tracks}
 \bibinfo{year}{[n.d.]}\natexlab{}.
\newblock \bibinfo{title}{Tracks}.
\newblock
\newblock
\newblock
\shownote{\url{https://github.com/TracksApp/tracks}.}


\bibitem[\protect\citeauthoryear{??}{ski}{[n.d.]}]%
        {skipvalidation}
 \bibinfo{year}{[n.d.]}\natexlab{}.
\newblock \bibinfo{title}{Validation API Document}.
\newblock
  \bibinfo{howpublished}{\url{https://guides.rubyonrails.org/v3.2/active_record_validations_callbacks.html\#skipping-callbacks}}.
\newblock


\bibitem[\protect\citeauthoryear{??}{hib}{[n.d.]}]%
        {hibernate-validation}
 \bibinfo{year}{[n.d.]}\natexlab{}.
\newblock \bibinfo{title}{Validators in Hibernate}.
\newblock \bibinfo{howpublished}{\url{https://hibernate.org/validator/}}.
\newblock


\bibitem[\protect\citeauthoryear{Abedjan, Schulze, and Naumann}{Abedjan
  et~al\mbox{.}}{2014}]%
        {abedjan2014dfd}
\bibfield{author}{\bibinfo{person}{Ziawasch Abedjan}, \bibinfo{person}{Patrick
  Schulze}, {and} \bibinfo{person}{Felix Naumann}.}
  \bibinfo{year}{2014}\natexlab{}.
\newblock \showarticletitle{DFD: Efficient functional dependency discovery}. In
  \bibinfo{booktitle}{\emph{Proceedings of the 23rd ACM International
  Conference on Conference on Information and Knowledge Management}}.
  \bibinfo{pages}{949--958}.
\newblock


\bibitem[\protect\citeauthoryear{Begoli, Camacho-Rodr{\'\i}guez, Hyde, Mior,
  and Lemire}{Begoli et~al\mbox{.}}{2018}]%
        {begoli2018apache}
\bibfield{author}{\bibinfo{person}{Edmon Begoli}, \bibinfo{person}{Jes{\'u}s
  Camacho-Rodr{\'\i}guez}, \bibinfo{person}{Julian Hyde},
  \bibinfo{person}{Michael~J Mior}, {and} \bibinfo{person}{Daniel Lemire}.}
  \bibinfo{year}{2018}\natexlab{}.
\newblock \showarticletitle{Apache calcite: A foundational framework for
  optimized query processing over heterogeneous data sources}. In
  \bibinfo{booktitle}{\emph{Proceedings of the 2018 International Conference on
  Management of Data}}. \bibinfo{pages}{221--230}.
\newblock


\bibitem[\protect\citeauthoryear{Brown and Haas}{Brown and Haas}{2003}]%
        {brown2003bhunt}
\bibfield{author}{\bibinfo{person}{Paul~G Brown} {and} \bibinfo{person}{Peter~J
  Haas}.} \bibinfo{year}{2003}\natexlab{}.
\newblock \showarticletitle{BHUNT: Automatic discovery of fuzzy algebraic
  constraints in relational data}. In \bibinfo{booktitle}{\emph{Proceedings
  2003 VLDB Conference}}. Elsevier, \bibinfo{pages}{668--679}.
\newblock


\bibitem[\protect\citeauthoryear{Chakravarthy, Grant, and Minker}{Chakravarthy
  et~al\mbox{.}}{1990}]%
        {rewrite-theory1}
\bibfield{author}{\bibinfo{person}{Upen~S. Chakravarthy}, \bibinfo{person}{John
  Grant}, {and} \bibinfo{person}{Jack Minker}.}
  \bibinfo{year}{1990}\natexlab{}.
\newblock \showarticletitle{Logic-Based Approach to Semantic Query
  Optimization}.
\newblock \bibinfo{journal}{\emph{ACM Trans. Database Syst.}}
  (\bibinfo{year}{1990}), \bibinfo{pages}{162–207}.
\newblock


\bibitem[\protect\citeauthoryear{Cheng, Gryz, Koo, Leung, Liu, Qian, and
  Schiefer}{Cheng et~al\mbox{.}}{1999}]%
        {cheng1999implementation}
\bibfield{author}{\bibinfo{person}{Qi Cheng}, \bibinfo{person}{Jarek Gryz},
  \bibinfo{person}{Fred Koo}, \bibinfo{person}{TY~Cliff Leung},
  \bibinfo{person}{Linqi Liu}, \bibinfo{person}{Xiaoyan Qian}, {and}
  \bibinfo{person}{Bernhard Schiefer}.} \bibinfo{year}{1999}\natexlab{}.
\newblock \showarticletitle{Implementation of two semantic query optimization
  techniques in DB2 universal database}. In \bibinfo{booktitle}{\emph{VLDB}},
  Vol.~\bibinfo{volume}{99}. Citeseer, \bibinfo{pages}{687--698}.
\newblock


\bibitem[\protect\citeauthoryear{Chu, Murphy, Roesch, Cheung, and Suciu}{Chu
  et~al\mbox{.}}{2018}]%
        {cosetteVLDB}
\bibfield{author}{\bibinfo{person}{Shumo Chu}, \bibinfo{person}{Brendan
  Murphy}, \bibinfo{person}{Jared Roesch}, \bibinfo{person}{Alvin Cheung},
  {and} \bibinfo{person}{Dan Suciu}.} \bibinfo{year}{2018}\natexlab{}.
\newblock \showarticletitle{Axiomatic Foundations and Algorithms for Deciding
  Semantic Equivalences of {SQL} Queries}.
\newblock \bibinfo{journal}{\emph{Proc. {VLDB} Endow.}} \bibinfo{volume}{11},
  \bibinfo{number}{11} (\bibinfo{year}{2018}), \bibinfo{pages}{1482--1495}.
\newblock


\bibitem[\protect\citeauthoryear{Cochrane, Pirahesh, and Mattos}{Cochrane
  et~al\mbox{.}}{1996}]%
        {cochrane1996integrating}
\bibfield{author}{\bibinfo{person}{Roberta Cochrane}, \bibinfo{person}{Hamid
  Pirahesh}, {and} \bibinfo{person}{Nelson Mattos}.}
  \bibinfo{year}{1996}\natexlab{}.
\newblock \showarticletitle{Integrating triggers and declarative constraints in
  SQL database systems}. In \bibinfo{booktitle}{\emph{VLDB}},
  Vol.~\bibinfo{volume}{96}. Citeseer, \bibinfo{pages}{3--6}.
\newblock


\bibitem[\protect\citeauthoryear{Dar, Franklin, J\'{o}nsson, Srivastava, and
  Tan}{Dar et~al\mbox{.}}{1996}]%
        {rewrite-system1}
\bibfield{author}{\bibinfo{person}{Shaul Dar}, \bibinfo{person}{Michael~J.
  Franklin}, \bibinfo{person}{Bj\"{o}rn~\TH{}\'{o}r J\'{o}nsson},
  \bibinfo{person}{Divesh Srivastava}, {and} \bibinfo{person}{Michael Tan}.}
  \bibinfo{year}{1996}\natexlab{}.
\newblock \showarticletitle{Semantic Data Caching and Replacement}. In
  \bibinfo{booktitle}{\emph{VLDB}}. \bibinfo{pages}{330–341}.
\newblock


\bibitem[\protect\citeauthoryear{De~Raedt, Passerini, and Teso}{De~Raedt
  et~al\mbox{.}}{2018}]%
        {de2018learning}
\bibfield{author}{\bibinfo{person}{Luc De~Raedt}, \bibinfo{person}{Andrea
  Passerini}, {and} \bibinfo{person}{Stefano Teso}.}
  \bibinfo{year}{2018}\natexlab{}.
\newblock \showarticletitle{Learning constraints from examples}. In
  \bibinfo{booktitle}{\emph{Thirty-Second AAAI Conference on Artificial
  Intelligence}}.
\newblock


\bibitem[\protect\citeauthoryear{Finance and Gardarin}{Finance and
  Gardarin}{1991}]%
        {finance1991rule}
\bibfield{author}{\bibinfo{person}{Beatrice Finance} {and}
  \bibinfo{person}{Georges Gardarin}.} \bibinfo{year}{1991}\natexlab{}.
\newblock \showarticletitle{A rule-based query rewriter in an extensible dbms}.
  In \bibinfo{booktitle}{\emph{Proceedings. Seventh International Conference on
  Data Engineering}}. IEEE Computer Society, \bibinfo{pages}{248--249}.
\newblock


\bibitem[\protect\citeauthoryear{Habimana}{Habimana}{2015}]%
        {habimana2015query}
\bibfield{author}{\bibinfo{person}{Jean Habimana}.}
  \bibinfo{year}{2015}\natexlab{}.
\newblock \showarticletitle{Query optimization techniques-tips for writing
  efficient and faster SQL queries}.
\newblock \bibinfo{journal}{\emph{International Journal of Scientific \&
  Technology Research}} \bibinfo{volume}{4}, \bibinfo{number}{10}
  (\bibinfo{year}{2015}), \bibinfo{pages}{22--26}.
\newblock


\bibitem[\protect\citeauthoryear{Huhtala, K{\"a}rkk{\"a}inen, Porkka, and
  Toivonen}{Huhtala et~al\mbox{.}}{1999}]%
        {huhtala1999tane}
\bibfield{author}{\bibinfo{person}{Yka Huhtala}, \bibinfo{person}{Juha
  K{\"a}rkk{\"a}inen}, \bibinfo{person}{Pasi Porkka}, {and}
  \bibinfo{person}{Hannu Toivonen}.} \bibinfo{year}{1999}\natexlab{}.
\newblock \showarticletitle{TANE: An efficient algorithm for discovering
  functional and approximate dependencies}.
\newblock \bibinfo{journal}{\emph{The computer journal}} \bibinfo{volume}{42},
  \bibinfo{number}{2} (\bibinfo{year}{1999}), \bibinfo{pages}{100--111}.
\newblock


\bibitem[\protect\citeauthoryear{Ilyas, Markl, Haas, Brown, and
  Aboulnaga}{Ilyas et~al\mbox{.}}{2004}]%
        {ilyas2004cords}
\bibfield{author}{\bibinfo{person}{Ihab~F Ilyas}, \bibinfo{person}{Volker
  Markl}, \bibinfo{person}{Peter Haas}, \bibinfo{person}{Paul Brown}, {and}
  \bibinfo{person}{Ashraf Aboulnaga}.} \bibinfo{year}{2004}\natexlab{}.
\newblock \showarticletitle{CORDS: Automatic discovery of correlations and soft
  functional dependencies}. In \bibinfo{booktitle}{\emph{Proceedings of the
  2004 ACM SIGMOD international conference on Management of data}}.
  \bibinfo{pages}{647--658}.
\newblock


\bibitem[\protect\citeauthoryear{Kimura, Huo, Rasin, Madden, and Zdonik}{Kimura
  et~al\mbox{.}}{2009}]%
        {kimura2009correlation}
\bibfield{author}{\bibinfo{person}{Hideaki Kimura}, \bibinfo{person}{George
  Huo}, \bibinfo{person}{Alexander Rasin}, \bibinfo{person}{Samuel Madden},
  {and} \bibinfo{person}{Stanley~B Zdonik}.} \bibinfo{year}{2009}\natexlab{}.
\newblock \showarticletitle{Correlation maps: a compressed access method for
  exploiting soft functional dependencies}.
\newblock \bibinfo{journal}{\emph{Proceedings of the VLDB Endowment}}
  \bibinfo{volume}{2}, \bibinfo{number}{1} (\bibinfo{year}{2009}),
  \bibinfo{pages}{1222--1233}.
\newblock


\bibitem[\protect\citeauthoryear{Levy and Sagiv}{Levy and Sagiv}{1995}]%
        {rewrite-theory2}
\bibfield{author}{\bibinfo{person}{Alon~Y. Levy} {and}
  \bibinfo{person}{Yehoshua Sagiv}.} \bibinfo{year}{1995}\natexlab{}.
\newblock \showarticletitle{Semantic Query Optimization in Datalog Programs
  (Extended Abstract)}. In \bibinfo{booktitle}{\emph{SIGMOD}}.
  \bibinfo{pages}{163–173}.
\newblock


\bibitem[\protect\citeauthoryear{Li, Wu, Yi, and Zhao}{Li
  et~al\mbox{.}}{2016}]%
        {li2016wander}
\bibfield{author}{\bibinfo{person}{Feifei Li}, \bibinfo{person}{Bin Wu},
  \bibinfo{person}{Ke Yi}, {and} \bibinfo{person}{Zhuoyue Zhao}.}
  \bibinfo{year}{2016}\natexlab{}.
\newblock \showarticletitle{Wander join: Online aggregation via random walks}.
  In \bibinfo{booktitle}{\emph{Proceedings of the 2016 International Conference
  on Management of Data}}. ACM, \bibinfo{pages}{615--629}.
\newblock


\bibitem[\protect\citeauthoryear{Liu, Xiao, Didwania, and Eltabakh}{Liu
  et~al\mbox{.}}{2016}]%
        {liu2016exploiting}
\bibfield{author}{\bibinfo{person}{Hai Liu}, \bibinfo{person}{Dongqing Xiao},
  \bibinfo{person}{Pankaj Didwania}, {and} \bibinfo{person}{Mohamed~Y
  Eltabakh}.} \bibinfo{year}{2016}\natexlab{}.
\newblock \showarticletitle{Exploiting soft and hard correlations in big data
  query optimization}.
\newblock \bibinfo{journal}{\emph{Proceedings of the VLDB Endowment}}
  \bibinfo{volume}{9}, \bibinfo{number}{12} (\bibinfo{year}{2016}),
  \bibinfo{pages}{1005--1016}.
\newblock


\bibitem[\protect\citeauthoryear{Meier, Schmidt, Wei, and Lausen}{Meier
  et~al\mbox{.}}{2013}]%
        {meier2013semantic}
\bibfield{author}{\bibinfo{person}{Michael Meier}, \bibinfo{person}{Michael
  Schmidt}, \bibinfo{person}{Fang Wei}, {and} \bibinfo{person}{Georg Lausen}.}
  \bibinfo{year}{2013}\natexlab{}.
\newblock \showarticletitle{Semantic query optimization in the presence of
  types}.
\newblock \bibinfo{journal}{\emph{J. Comput. System Sci.}}
  \bibinfo{volume}{79}, \bibinfo{number}{6} (\bibinfo{year}{2013}),
  \bibinfo{pages}{937--957}.
\newblock


\bibitem[\protect\citeauthoryear{Minker}{Minker}{1988}]%
        {rewrite-theory0}
\bibfield{editor}{\bibinfo{person}{J. Minker}} (Ed.).
  \bibinfo{year}{1988}\natexlab{}.
\newblock \bibinfo{booktitle}{\emph{Foundations of Deductive Databases and
  Logic Programming}}.
\newblock \bibinfo{publisher}{Morgan Kaufmann Publishers Inc.}
\newblock


\bibitem[\protect\citeauthoryear{Novelli and Cicchetti}{Novelli and
  Cicchetti}{2001}]%
        {novelli2001fun}
\bibfield{author}{\bibinfo{person}{Noel Novelli} {and} \bibinfo{person}{Rosine
  Cicchetti}.} \bibinfo{year}{2001}\natexlab{}.
\newblock \showarticletitle{Fun: An efficient algorithm for mining functional
  and embedded dependencies}. In \bibinfo{booktitle}{\emph{International
  Conference on Database Theory}}. Springer, \bibinfo{pages}{189--203}.
\newblock


\bibitem[\protect\citeauthoryear{Papenbrock and Naumann}{Papenbrock and
  Naumann}{2016}]%
        {papenbrock2016hybrid}
\bibfield{author}{\bibinfo{person}{Thorsten Papenbrock} {and}
  \bibinfo{person}{Felix Naumann}.} \bibinfo{year}{2016}\natexlab{}.
\newblock \showarticletitle{A hybrid approach to functional dependency
  discovery}. In \bibinfo{booktitle}{\emph{Proceedings of the 2016
  International Conference on Management of Data}}. \bibinfo{pages}{821--833}.
\newblock


\bibitem[\protect\citeauthoryear{Papenbrock and Naumann}{Papenbrock and
  Naumann}{2017}]%
        {papenbrock2017hybrid}
\bibfield{author}{\bibinfo{person}{Thorsten Papenbrock} {and}
  \bibinfo{person}{Felix Naumann}.} \bibinfo{year}{2017}\natexlab{}.
\newblock \showarticletitle{A hybrid approach for efficient unique column
  combination discovery}.
\newblock \bibinfo{journal}{\emph{Datenbanksysteme f{\"u}r Business,
  Technologie und Web (BTW 2017)}} (\bibinfo{year}{2017}).
\newblock


\bibitem[\protect\citeauthoryear{Paulley}{Paulley}{2001}]%
        {FD-QO-theory}
\bibfield{author}{\bibinfo{person}{Glenn~Norman Paulley}.}
  \bibinfo{year}{2001}\natexlab{}.
\newblock \bibinfo{booktitle}{\emph{Exploiting functional dependence in query
  optimization}}.
\newblock \bibinfo{publisher}{Citeseer}.
\newblock


\bibitem[\protect\citeauthoryear{Pirahesh, Hellerstein, and Hasan}{Pirahesh
  et~al\mbox{.}}{1992}]%
        {pirahesh1992extensible}
\bibfield{author}{\bibinfo{person}{Hamid Pirahesh}, \bibinfo{person}{Joseph~M
  Hellerstein}, {and} \bibinfo{person}{Waqar Hasan}.}
  \bibinfo{year}{1992}\natexlab{}.
\newblock \showarticletitle{Extensible/rule based query rewrite optimization in
  Starburst}.
\newblock \bibinfo{journal}{\emph{ACM Sigmod Record}} \bibinfo{volume}{21},
  \bibinfo{number}{2} (\bibinfo{year}{1992}), \bibinfo{pages}{39--48}.
\newblock


\bibitem[\protect\citeauthoryear{Ramakrishnan and Gehrke}{Ramakrishnan and
  Gehrke}{2000}]%
        {bcnf}
\bibfield{author}{\bibinfo{person}{Raghu Ramakrishnan} {and}
  \bibinfo{person}{Johannes Gehrke}.} \bibinfo{year}{2000}\natexlab{}.
\newblock \bibinfo{booktitle}{\emph{Database management systems}}.
\newblock \bibinfo{publisher}{McGraw-Hill}.
\newblock


\bibitem[\protect\citeauthoryear{Shenoy and Ozsoyoglu}{Shenoy and
  Ozsoyoglu}{1987}]%
        {rewrite-theory4}
\bibfield{author}{\bibinfo{person}{Sreekumar~T. Shenoy} {and}
  \bibinfo{person}{Z.~Meral Ozsoyoglu}.} \bibinfo{year}{1987}\natexlab{}.
\newblock \showarticletitle{A System for Semantic Query Optimization}. In
  \bibinfo{booktitle}{\emph{SIGMOD}}. \bibinfo{pages}{181–195}.
\newblock


\bibitem[\protect\citeauthoryear{Yan, Schulte, Zhang, Wang, and Cheng}{Yan
  et~al\mbox{.}}{2020}]%
        {jiannan:sigmod20:error-detection}
\bibfield{author}{\bibinfo{person}{Jing~Nathan Yan}, \bibinfo{person}{Oliver
  Schulte}, \bibinfo{person}{MoHan Zhang}, \bibinfo{person}{Jiannan Wang},
  {and} \bibinfo{person}{Reynold Cheng}.} \bibinfo{year}{2020}\natexlab{}.
\newblock \showarticletitle{SCODED: Statistical Constraint Oriented Data Error
  Detection}. In \bibinfo{booktitle}{\emph{Proceedings of the 2020 ACM SIGMOD
  International Conference on Management of Data}}. \bibinfo{pages}{845--860}.
\newblock


\bibitem[\protect\citeauthoryear{Yang, Sethi, Yan, Lu, and Cheung}{Yang
  et~al\mbox{.}}{2020}]%
        {yang:icse20}
\bibfield{author}{\bibinfo{person}{Junwen Yang}, \bibinfo{person}{Utsav Sethi},
  \bibinfo{person}{Cong Yan}, \bibinfo{person}{Shan Lu}, {and}
  \bibinfo{person}{Alvin Cheung}.} \bibinfo{year}{2020}\natexlab{}.
\newblock \showarticletitle{Managing Data Constraints in Database-Backed Web
  Applications}. In \bibinfo{booktitle}{\emph{ICSE}}.
\newblock


\bibitem[\protect\citeauthoryear{Yang, Yan, Subramaniam, Lu, and Cheung}{Yang
  et~al\mbox{.}}{2018}]%
        {yang2018not}
\bibfield{author}{\bibinfo{person}{Junwen Yang}, \bibinfo{person}{Cong Yan},
  \bibinfo{person}{Pranav Subramaniam}, \bibinfo{person}{Shan Lu}, {and}
  \bibinfo{person}{Alvin Cheung}.} \bibinfo{year}{2018}\natexlab{}.
\newblock \showarticletitle{How not to structure your database-backed web
  applications: a study of performance bugs in the wild}. In
  \bibinfo{booktitle}{\emph{2018 IEEE/ACM 40th International Conference on
  Software Engineering (ICSE)}}. IEEE, \bibinfo{pages}{800--810}.
\newblock


\bibitem[\protect\citeauthoryear{Yao and Hamilton}{Yao and Hamilton}{2008}]%
        {yao2008mining}
\bibfield{author}{\bibinfo{person}{Hong Yao} {and} \bibinfo{person}{Howard~J
  Hamilton}.} \bibinfo{year}{2008}\natexlab{}.
\newblock \showarticletitle{Mining functional dependencies from data}.
\newblock \bibinfo{journal}{\emph{Data Mining and Knowledge Discovery}}
  \bibinfo{volume}{16}, \bibinfo{number}{2} (\bibinfo{year}{2008}),
  \bibinfo{pages}{197--219}.
\newblock


\bibitem[\protect\citeauthoryear{{Zuohao She}, {Ravishankar}, and
  {Duggan}}{{Zuohao She} et~al\mbox{.}}{2016}]%
        {rewrite-system2}
\bibfield{author}{\bibinfo{person}{{Zuohao She}}, \bibinfo{person}{S.
  {Ravishankar}}, {and} \bibinfo{person}{J. {Duggan}}.}
  \bibinfo{year}{2016}\natexlab{}.
\newblock \showarticletitle{BigDAWG polystore query optimization through
  semantic equivalences}. In \bibinfo{booktitle}{\emph{2016 IEEE High
  Performance Extreme Computing Conference (HPEC)}}. \bibinfo{pages}{1--6}.
\newblock


\end{thebibliography}

\end{document}